# Scalable fabrication of gap-plasmon-based dynamic and chromogenic nanostructures by capillary-interaction driven self-assembly of liquid-metal

Renu Raman Sahu, Alwar Samy Ramasamy, Santosh Bhonsle, Mark Vailshery, Tapajyoti Das Gupta*

Laboratory of Advanced Nanostructures for Photonics and Electronics, Department of Instrumentation and Applied Physics, Indian Institute of Science, C. V. Raman Road, Bengaluru-560012, India

E-mail: tapajyoti@iisc.ac.in

Dynamically tunable nanoengineered structures for coloration show promising applications in sensing, displays, and communication. However, their potential challenge remains in having a scalable manufacturing process over large scales in 10s of cm of area. For the first time, we report a novel approach for fabricating chromogenic nanostructures that respond to mechanical stimuli by utilizing the fluidic properties of polydimethylsiloxane (PDMS) as a substrate and the interfacial tension of liquid metal-based plasmonic nanoparticles. Relying on the PDMS tunable property and a physical deposition method, our approach is single-step, scalable, and does not rely on high carbon footprint lithographic processes. By tuning the oligomer content in PDMS, we show that varieties of structural colors covering a significant gamut in CIE coordinates are achieved. We develop a model which depicts the formation of Ga nanodroplets from the capillary interaction of oligomers in PDMS with Ga. We showcase the capabilities of our processing technique by presenting prototypes of reflective displays and sensors for monitoring body parts, smart bandages, and the capacity of the nanostructured film to map force in real-time. These examples illustrate this technology's broad range of applications, such as large-area displays, devices for human-computer interactions, healthcare, and visual communication.

**Introduction**

Structurally colored materials bestowed with chromatic responsiveness[1–3] to external stimuli find enormous utility in visual communication[4] and sensing[5] applications. Despite their usefulness, the numerous manufacturing steps make it difficult to scale up, which presents a significant obstacle to their widespread use in industrial and household settings.[6,7].

Several chromogenic structures have been fabricated with mechanoresponsive properties[8]. While the existing top-down approaches for forming chromogenic structures offer precise structural control and reliability, they are not viable for fabrication on elastomeric substrates[9]. Such techniques employ lithographic steps and are primarily applied to fabricate structures on rigid substrates like silicon[10] and silicon nitride[11,12]. Mechanical responsiveness is incorporated by transferring the pattern to elastomeric substrates [13] or by laser printing[14] on dielectric-coated elastomeric substrates[15] making the process less scalable. Other methods include using a photosensitive polymer with a periodic refractive index variation [16, 17], nanoimprinting[18,19], or a combination of lithographic steps[20–22]. However, they still require lithographically fabricated masters and are not amenable to large-area fabrication unless proper alignment of the master is provided. On the other hand, most of the bottom-up approaches, usually involving self-assembly,[23–25] have been demonstrated to exhibit structural colors. Such techniques involve multiple processing steps and complex chemical methods and are thus not suitable for a cost-effective fabrication with high fidelity. These limitations necessitate a scalable fabrication technique with a single process step that allows control of the nanoscale morphology with precise control of the dynamically tunable structural color.

The use of liquid metals, such as Gallium [26–29], and its alloys [30–33] allow for the fabrication of flexible, bendable, and mechanically reconfigurable devices when deposited on an elastomeric substrate [34,35]. They have been the choice materials for electronic applications, particularly in the domain of soft and stretchable electronics[36]. Thus far, however, their application in the visible photonics domain remains unexplored due to its inability to be scaled down to nanoscale owing to their intrinsic large surface tension of 708 mN/m at room temperature[28], and thus preventing fabrication of dynamic photonic systems with sub 100 nm feature dimensions and its application in the visible domain of the optical spectrum.

In this work, we overcome this challenge and control the features in these length scales by the capillary interaction of nanodroplets of Gallium with the PDMS substrate. Thermally cured PDMS has uncrosslinked low molecular weight chains of PDMS called oligomers, which remain in a liquid state, infused in its crosslinked matrix [37,38]. By exploiting the large contrasting difference in the surface tensions of Gallium and liquid oligomers, we attain nanodroplets of Gallium with the substrate filling the gap between each droplet. The plasmonic properties of the Gallium nanodroplets on the elastomer produce mechanically responsive structural colors that cover a significant portion of chromaticity coordinates with excellent reproducibility. Our approach is a single-step process capable of fabricating the structural color on a large area; thus, a fast, scalable, and affordable technology. Further, we introduce a mathematical model that confirms our experimental findings and allows for the determination of the nanostructure of Ga through its fluidic interaction with the liquid oligomers of PDMS. The resulting structural and optical properties and the chromatic tunability of the structure are characterized.

To show our scalable process, we fabricate palletes of structural colors of the films derived from this process, which can be used in reflective displays. The optical response of the sample with respect to a minimum of 1000 periodic cycles of strain is shown to be repeatable, thus proving the reliability and robustness of the sample response to the periodic mechanical strain. We demonstrate the prototypes of healthcare monitoring devices like a smart bandage, body-part monitor, visual strain gauge for healthcare workers, and colorimetric strain sensor with a computer interface that takes input from the sample to give real-time signals. Our fabricated structure has further potential to be applied in advanced human-computer interfaces, soft robotics, and prosthetics.

**Fabrication of Ga-based nanostructures for coloration**

The fabrication approach we employ is depicted in Figure 1a and detailed in the method and SI section (SI Figures 1a, 1b and 1c). PDMS ratio X is made by proportionate mixing of X g of base and 1 g of curing agent (Sylgard 184, Dow Corning) during the substrate preparation. The proportion of the curing agent in the PDMS determines the extent of crosslinking and hence the softness of the substrate (SI Figure 1b). Moreover, it also determines the volume of oligomers in the crosslinked PDMS matrix[39,40], which is crucial in determining the final optical properties.

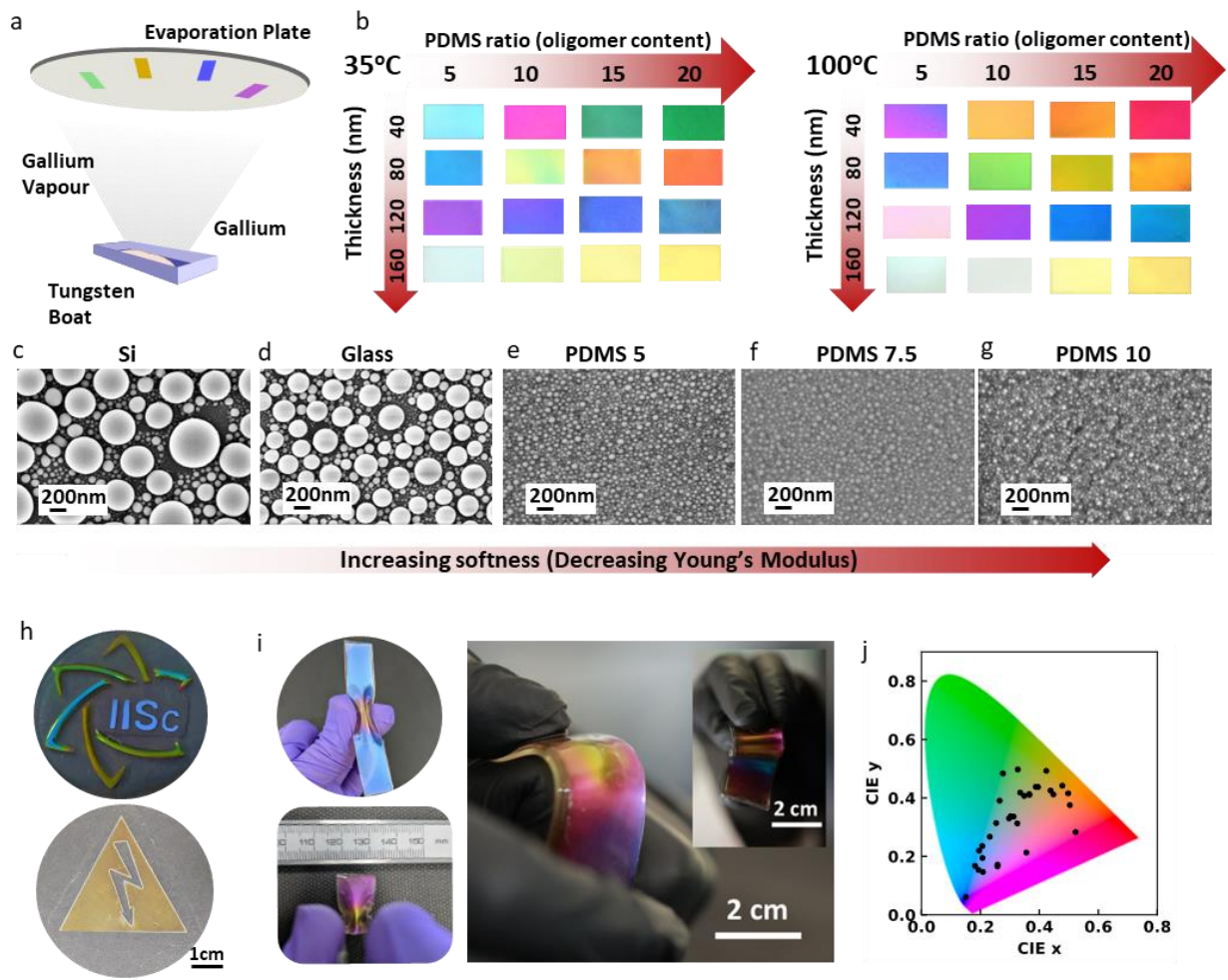

*Figure 1 | a, Thermal evaporation of Gallium onto the substrates. b, Color pallete obtained by tuning the PDMS oligomer content, with Ga deposited at different deposition parameters. The color represented here are the cropped images of the Ga-deposited samples. c-g, Top view scanning electron microscopy images of Ga deposited onto Silicon, Glass, PDMS 5, PDMS 7.5, and PDMS 10. The scale bar represents 200nm. h, Chromogenic structures on large areas (30 cm$^2$) for IISc logo and symbols. i, Mechanoresponsiveness of structural colors. j, Chromaticity coordinates of the fabricated samples on the CIE diagram.*

We observe a distinct coloration (Figure 1b) as we change the PDMS ratio (from PDMS 5 to 20), substrate temperature and the Gallium deposited's thickness. Furthermore, by tuning the substrate softness (Silicon (Si), Glass, and varying proportion of PDMS (Figure 1c-g)), we observe different particle size distributions of Ga nanoparticles on different substrates (Figure 1c-g and SI Figure 1d-e). On typical rigid substrates like Si (Youngs Modulus ~150 GPa)[38] and Glass (Youngs Modulus ~60 GPa)[41], the top view SEM images (Figure 1c, 1d) show large size distribution of Ga droplets around 70nm and 40 nm respectively, which are poly-dispersed. While in PDMS substrates (Figure 1e-g and SI Figure 1d) (Young's modulus ranging from 0.3 MPa to 1.3MPa, SI Figure 2), droplets are narrowly dispersed (radii with a mean and standard deviation of around 20nm and 10nm respectively) as depicted in the histogram plot (SI Figure 1e). The average sizes of the Ga nanodroplets as well as their dispersity, decrease as we increase the substrate softness (Figure 1c-g and SI Figure 2), indicating a positive correlation between the nanostructure's sizes and the elastic modulus of the substrate.

The elastic modulus of PDMS substrates varies from 0.3 to 1.3 MPa, much smaller than typical rigid substrates. Despite this, there are significant variations in color, prompting an investigation into other factors that impact the formation of Ga nanostructures in PDMS beyond just the softness of the substrate. Furthermore, the wide variation of color despite the slight change in size distribution observed from top-view SEM images (Figure 1c-g), as shown in Figure 1b, necessitates a detailed cross-sectional morphology of the Ga nanostructure. At the foresight, one can infer from the processing of the PDMS substrate that the oligomers remaining in the PDMS substrate participate in the nanostructure formation more prominently than the elastic modulus of the substrate does, as will be explained in the following sections. Having identified the processing step of mixing the PDMS base with the curing agent determines the softness and oligomer content in the substrate, the fabrication approach is simple and scalable. As evident from figure 1h, large areas order of a few cm sq. can be fabricated, such as the IISc logo structurally colored over 30 cm$^2$, an area similar to those of petri-dishes. The PDMS substrate, being flexible, allows for dynamic tuning of the structural color via mechanical deformation (Figure 1i). The wide array of structural colors, thus fabricated, spans a significant fraction of chromaticity coordinates (Figure 1j) by leveraging on two key factors: (1) substrate softness and, more prominently, (2) oligomer content in PDMS.

**Role of substrate properties: Softness and Oligomer content**

The interaction between capillary and elastic forces in the nanoscale is more significant in soft substrates than hard substrates[42]. The micron or nano-sized droplets on soft substrates form a dimple on the substrate due to Laplace pressure inside the droplet and the surface tension force at the three-phase contact line of the droplet pulls the substrate (Figure 2a)[43–45]. PDMS substrate deposited with Ga is sliced with a Focused Ion Beam (FIB) and imaged in the cross-section with a Scanning electron microscope to study the interactions of Ga with the substrate. The cross-sectional SEM image of PDMS 5 and 20 reveals the multi-layers of Ga nanodroplets beneath the top surface rather than forming a single layer with deformation on the substrate (Figure 2b left). We observe a distinct increase in the number of layers and a decrease in particle size as we increase the PDMS ratio, thus attributing primarily to the volume conservation of deposited Gallium. Furthermore, the radius of Ga nanodroplets decreases with an increase in depth into the PDMS substrate (Figure 2b left).

Although the oligomers do not play a significant role in determining the elastic modulus of the substrate (SI Figure 2), their presence largely determines the Ga nanostructures that form upon deposition on the substrate. On removal of the oligomers from PDMS by Toluene treatment (see methods), we observe a monolayer of Gallium nanodroplets with dimple formation on the substrate (Figure 2b right and SI Figure 3). In addition, removing the oligomers eliminates the difference in the size distribution of Ga nanodroplets on different PDMS ratios (Figure 2b right and SI Figure 4). Consequently, the mixing ratios of the PDMS base to the crosslinker are of little significance once the oligomers are removed. And the resulting colors resemble white (Figure 2b inset and SI Figure 5) as observed in Glass and Si substrates. Thus, the importance of oligomers in PDMS to form distinct structural colors, as shown in Figure 1a, is substantiated.

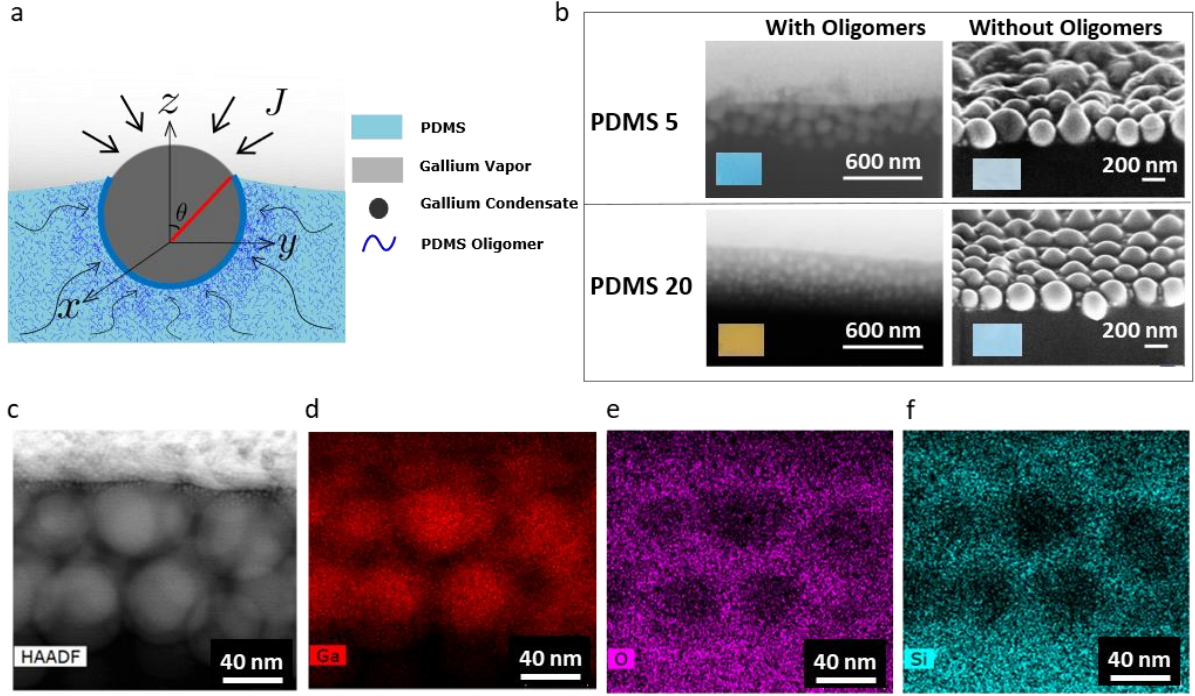

*Figure 2 | a, Schematic diagram of Ga nanodroplet on PDMS extracting out the oligomers. b, Cross-section SEM images of PDMS 5 and 20, with and without oligomers. The inset shows the image of the fabricated sample. c, High-Angle Annular Dark Field Image of the cross-section of Ga deposited on PDMS 5. d, e, f, Energy Dispersive X-ray map of Ga, Si and O respectively.*

**The hypothesis of the formation of multi-layers of Ga nanodroplets in PDMS**

To gain a quantitative understanding of Ga nanodroplets formed on PDMS, we propose a hypothesis based on the spreading of oligomers to minimize the surface energy of the system. The spreading of one fluid occurs over the other to minimize the surface energy, which occurs when the spreading parameter is positive. The spreading parameter is defined as,

$$S = \gamma_{Ga} - (\gamma_{oligomers-Ga} + \gamma_{oligomers}) \qquad (1)$$

where $\gamma_{Ga}$ and $\gamma_{oligomers}$ are the surface tension of the Ga and oligomers respectively and $\gamma_{oligomers-Ga}$ is the interfacial tension between the Ga and oligomers. For the configuration of the Ga droplet on PDMS, interfacial tension $\gamma_{Ga}$, $\gamma_{oligomers}$ and $\gamma_{oligomers-Ga}$ are 650, 20 and 590 mN/m respectively (measured through the Pendant drop method with Goniometer, Dataphysics OCA 25). The spreading coefficient is calculated to be 40 mN/m, for which the oligomers present in the PDMS can form a cloaking layer around the Ga droplets. A cloaking layer of the oligomers around the Ga nanodroplet is confirmed by High-Angle Annular Dark Field (HAADF) image (Figure 2c), where one can distinctly observe the separation between the droplets. The Energy Dispersive X-ray (EDAX) map of Ga reveals the spatial distribution of droplets (Figure 2d). The EDAX map of Silicon (Figure 2e) and Oxygen (Figure 2f) indicates the presence of PDMS oligomers between the droplets, corroborating that the gap filled with dielectric. The next layer of Ga vapor will form on top of the 1st layer of droplets since the oligomers resist the coalescence of Ga nanodroplets, as confirmed by the HAADF image and EDAX maps. By the mechanism mentioned above, the oligomers will encapsulate the next layer as well; however, comparatively slower than that of the previous layer because of their less availability. This results in slower encapsulation of the second layer of droplets, allowing a larger window of time to grow in size due to the condensation of Ga vapor. This mechanism repeats for the further layers until the oligomer content is insufficient for encapsulation. Hence, the Ga nanodroplets exhibit a decreasing radii trend with an increase in the depth of the layer, confirming the experimental observation (Figure 2b left).

**Modeling: Ga nanodroplet layer formation on PDMS substrates**

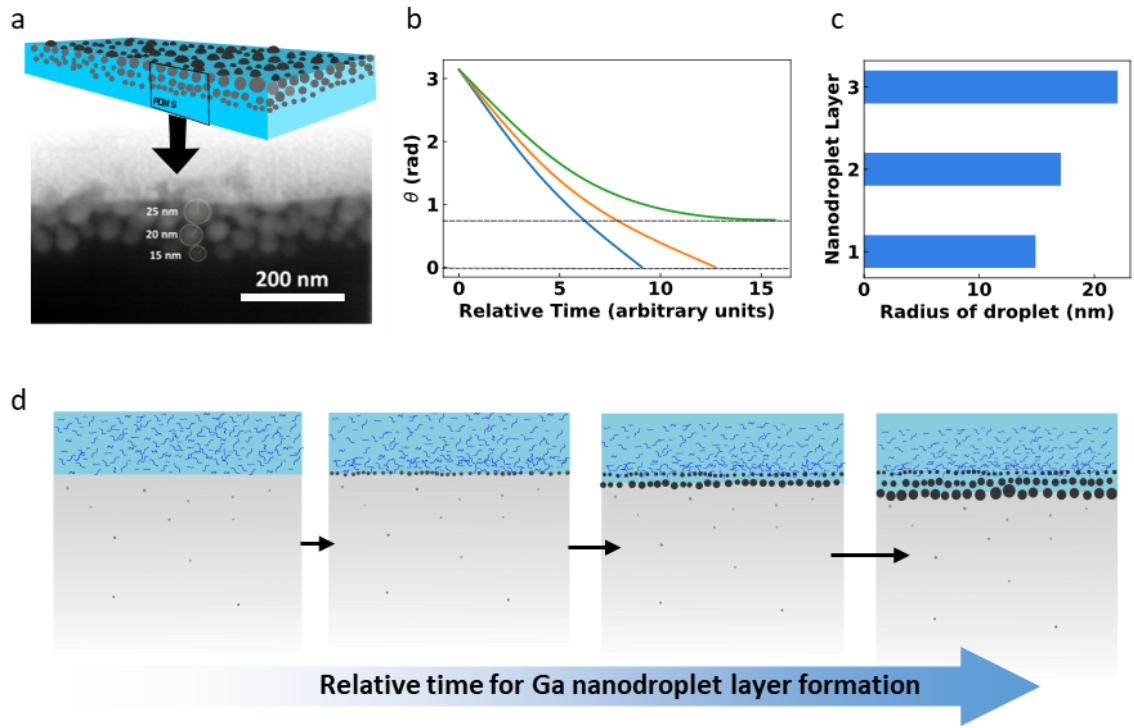

*Figure 3 | a, Schematic cross-section view of Ga deposited on PDMS and the cross-section SEM image from a thin slice cut with FIB. b, Relative time scale of engulfing of the Ga nanodroplet layers. c, Radii of Ga nanodroplets as predicted from the SGE equations. d, Schematic depiction of the formation of layered nanodroplets when Ga is thermally evaporated onto PDMS. The length of the black arrows depicts the relative time it takes to form one layer to the next.*

Based on the hypothesis proposed above, we developed a model to determine the radius of the Ga droplets in different layers and the relative time scale in layer formation for a given PDMS ratio and deposition parameters assuming the droplet to be spherical (SI Figure 6 and 7). The schematic diagram of the structure and the corresponding cross-section SEM image is shown in Figure 3a, depicting the radii trend with an increase in the depth into the substrate.

Thermal evaporation leads to Gallium vapors that condense as it approaches the substrate and form spherical condensates to minimize their surface energies (SI section 2.1). Once these condensates, which are in a liquid state, reach the substrate, there would be two phenomena happening: (1) Deformation of the substrate owing to the Laplace pressure of the droplet (SI section 2.2), and (2) Extraction of the liquid oligomers to engulf the layer of condensate droplets (SI section 2.3, 2.4). For the engulfing to happen, there must be a copious volume of oligomers available at the interface of Ga droplet and PDMS, leading us to the assumption that the rate of engulfing is proportional to the volume of oligomers. This process is quantified by the rate of change of the engulfing angle, defined as the angle between the normal to the PDMS substrate and the radial line from the center of the droplet to the interphase circle (SI Figure 8 and 9). The increase of radius with respect to time, which occurs due to the condensation of vaporized Ga atoms into the liquid droplet (SI Figure 10), quantifies the growth of the nanodroplet. The time scales of engulfing and growth of nanodroplets during the deposition determine the sizes of Gallium nanodroplets and the number of layers formed along the depth of the PDMS (SI Figure 11-13). The interplay of co-occurring events are described in SI section 2.5. The migration of oligomers to PDMS-Gallium interface (SI section 2.6), growth of Ga nanodroplet during condensation (SI section 2.7), and its engulfing (SI section 2.8) are described by the set of following three equations respectively,

$$\frac{df}{dt} = -k_s\big(f - 2\pi R^2 w(1 + \cos\theta)\big)(2\pi R \sin\theta) \quad (1)$$

$$\frac{dR}{dt} = k_\rho(1 - \cos\theta) \quad (2)$$

$$\frac{d\theta}{dt} = -k_e\big(f - 2\pi R^2 w(1 + \cos\theta)\big) \quad (3)$$

where, $f$ is the volume of liquid oligomers present in a given substrate volume, in units of nm³, $R$ is the radius of the Gallium droplet, and $\theta$ is the engulfing angle. The flux of Gallium atoms condensing onto the droplets is given by $J$, and $w$ is a thin layer of oligomers encapsulating the nanodroplet during engulfing. The constants $\kappa_s$, $\kappa_\rho$ and $\kappa_s$ are constants of proportionality for equations 2, 3 and 4 respectively (SI sections 2.5-2.8 for details).

These equations describe the simultaneous occurrence of three events for a single Ga nanodroplet, namely, (i) the availability of oligomers at the Ga-PDMS interface, (ii) the growth of Ga nanodroplet during the process of its engulfing, (iii) engulfing of the nanodroplet by liquid oligomers (Figure 2a). Thus, equations 2, 3, and 4 are termed the substrate (S), growth(G) and engulfing(E) equations, respectively. They are solved iteratively (see SI section 2.9, Iterative procedure to obtain the number of immersed layers) for each layer. Figure 3b depicts the relative time scale of immersion of the first and second layer of the Ga nanodroplets, clearly matching the explanation mentioned above in the previous section that the first (innermost) layer takes the least time to form and the time of formation of the subsequent layers increases. The third layer, in this case, is partially immersed due to insufficient availability of oligomers for engulfing, thus terminating the process for further engulfing (Figure 3b). The final sizes of the droplets in a given layer are determined by the available time window for growth before the engulfment of the droplets. During this time window, as depicted in the x-axis of Figure 3b, the Ga nanodroplets grow in size from the critical radius at which they condense to the values shown in Figure 3c (SI section 2.10 and 2.11). Figure 3d is a schematic depiction of the formation of different layers of Ga nanodroplets on a relative timescale. The parameters $w$, and the SGE constants $\kappa_s$, $\kappa_\rho$ and $\kappa_e$ are determined by a fitting method for a given PDMS substrate and deposition parameters (SI Figure 15). The parameter $\kappa_\rho$ depends on the deposition parameters while $\kappa_e$ and $\kappa_s$ depends on the substrate-liquid metal interaction. For a given set of parameters for deposition and the substrate, one can obtain the constants $\kappa_s$, $\kappa_e$ and $\kappa_\rho$ (SI Figure 15). These equations determine the number of layers of Ga droplets that will be formed on PDMS and the size trend of the radii of the deposited nanodroplet with oligomer content, rate, and deposition thickness. This analysis clearly emphasizes that a suitable choice of thickness, rate, PDMS oligomer content, and substrate temperature leads to various structural colorations. To further explore the validity of our proposed SGE equations, we investigated the effect of the deposition rate (SI Figure 16), keeping substrate and all other deposition parameters the same. The SEM images (SI Figure 17) reveal the trends that match those from the SGE equations; as the rate of deposition increases, the radii of Gallium nanodroplets increase as well, while the number of layers formed decreases. Thus, these equations can be of immense applicability while designing process steps to obtain a desired size of Ga nanodroplets and layers.

**Spectral Trend caused by Oligomer content in the substrate**

The resultant morphology of Ga nanodroplets on the PDMS substrates obtained by tuning the oligomer content influences the spectral properties of the samples. The experimentally obtained reflectivity spectra red-shifts, owing to the structural variation of the Ga nanodroplets, as the PDMS ratio increases (Figure 4a). The structure was simulated in Lumerical, commercial software for solving Maxwell's equations using the Finite Difference Time Domain (FDTD) numerical method. The optical properties of Ga are determined by its wavelength-dependent complex refractive index (SI Figure 18). The spectra obtained from the simulation match the experimental trend (Figure 4b, SI section 3.1, SI Figure 19).

A minimum wavelength of light exists for a given structure characterized by a certain number of layers, which can excite the fields between the inter-layer and intra-layer Ga nanodroplets of all the layers (see SI figure 20). The highest peak in the reflectivity spectrum corresponds to the wavelength that allows the field to interact with all layers of the structure (Figure 4c). One might expect a blue shift due to the reduction in the size of Ga nanodroplets with increasing oligomer content, but instead, we observed a red shift in the reflectivity spectra (Figure 4d). This is because the thickness of the layer that contains the Ga nanodroplets tends to increase as we adjust the oligomer content (refer to SI Figure 21 and 22). Thus, we conclude that the collective effect from all the layers plays a role in determining the spectral features from the sample.

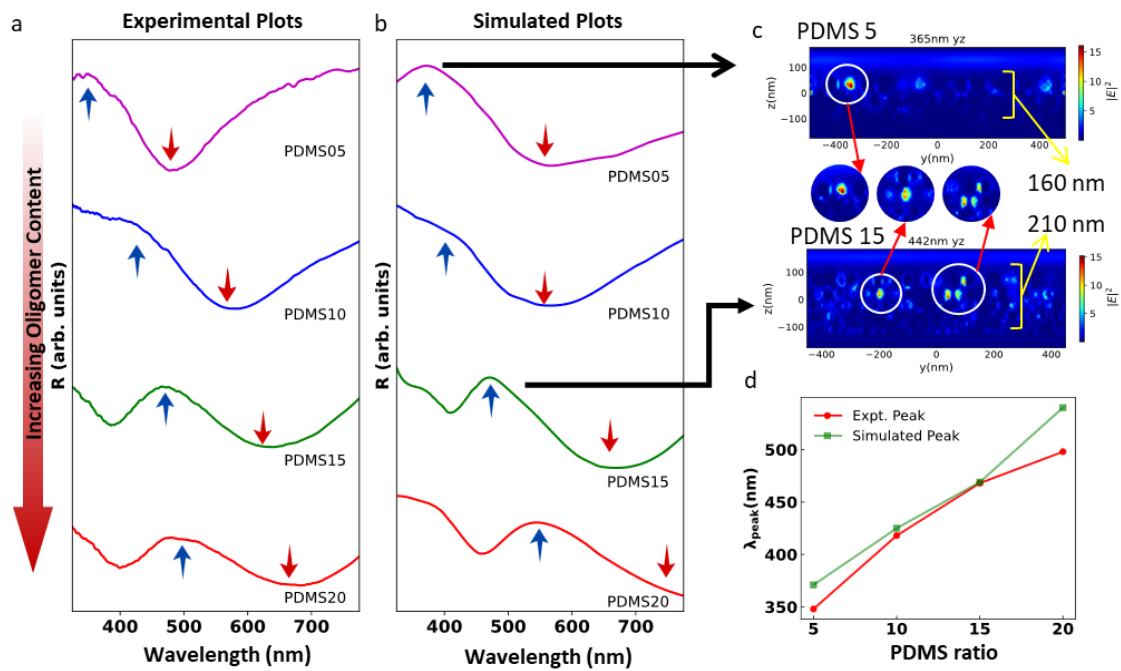

*Figure 4 | Spectral trend with oligomer content |a, Reflectivity Spectrum obtained experimentally. b, Reflectivity spectrum from FDTD simulations. c, The Electric Field Intensity of simulated PDMS 5 structures at 365nm and PDMS 15 at 442 nm. d, Wavelength at which the reflectivity peaks vs. PDMS ratios.*

## Spectral trend caused by mechanical strain: Dynamic tuning of the optical properties

The mechanical deformation due to stretching will significantly affect the interparticle gap along the stretching direction. Accordingly, we examined the interaction between the Gallium nanoparticles in a single layer to understand the optical properties caused by the Gallium nanodroplets on stretching. The local hotspots of the electric field intensity plots (circularly zoomed regions in Figure 4c) are the plasmonic resonances from the gap between two nearly placed Ga nanodroplets. When the whole structure is excited, there is a significant contribution from all the intra and inter-layer gap-plasmons. These hotspots play a crucial role in determining the spectra in the visible region, as explained subsequently. Experimentally on applying a uniaxial strain, a change in the color of the sample (Figure 5a) is observed due to a blue shift of the reflectivity spectra (Figure 5b, SI videos 1 and 2). The spectral features shift by 3.68nm per mm of change in sample length (Figure 5c). With the hue of the sample changing from 6 to 65 degrees (Figure 5d), the change in color as depicted in the chromaticity coordinates (Figure 5e) reveals a discernible and significant shift on uniaxial stretching. The spectral change is significantly repeatable and reversible over a minimum of 1000 cycles of the applied periodic strain (Figure 5f) of at least 60% (SI Figure 23, SI section 3.2, SI video 3), thus proving it to be a reliable chromatic sensor with high longevity.

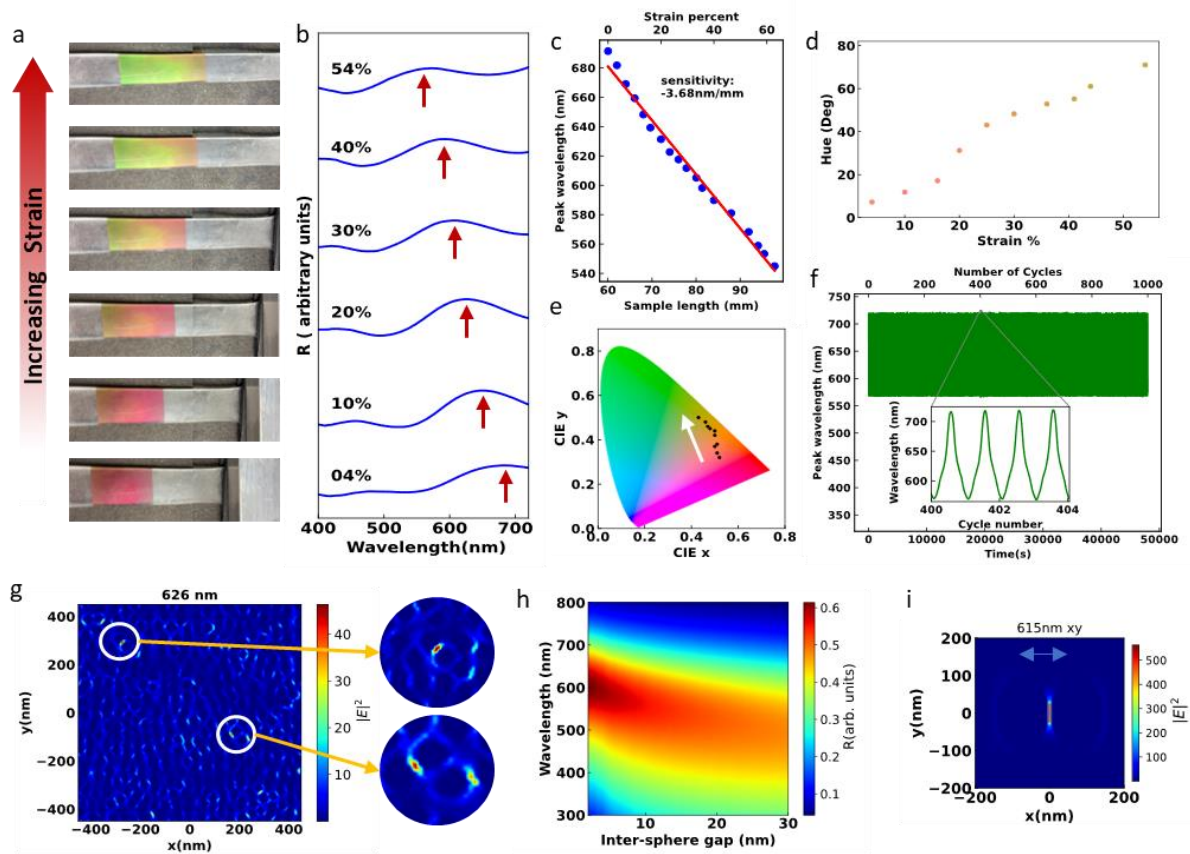

*Figure 5 | Uniaxial strain on Ga deposited PDMS. a, Color of the sample from pink to green. b, The blue shift of the reflectivity spectrum on applying linear uniaxial strain. c, The peak wavelength of reflectivity vs. length of the sample and strain (%). d, Change of hue with respect to applied strain. e, Change of Chromaticity coordinates with strain. f, Peak reflectivity vs. time of a periodic uniaxial linear strain for 1000 cycles. g, Electric field intensity plot across one of the layers of simulated structure. The circularly zoomed regions depict the occurrence of gap plasmon resonances. h, Simulation of reflectivity from a dimer structure of two spheres of radius 50nm placed near each other with a varying gap. i, Enhanced field intensity between two Ga nanospheres with a 2nm surface-to-surface gap in between them.*

The spectral contribution due to the gap plasmon resonances occurs due to plasmonic excitation in the dimeric pairs, as depicted in the field intensity plot of a particular layer of the simulated structure

(Figure 5g). Applying a uniaxial strain increases the gap between the droplets placed near each other. One observes a blue shift of the simulated reflectivity spectra (Figure 5h), with the increasing dimer gap and the polarisation vector oriented toward the gap (Figure 5i, SI section 3.3 and 3.4 and SI Figure 24). Since, in the experimental conditions, the Ga nanodroplets are distributed randomly in the plane of the layer, unpolarised light can excite the plasmonic hotspots. Apart from the gap plasmons, which determine the dynamic behaviour of the PDMS substrate on mechanical deformation, the size of the Gallium nanodroplets plays a significant role in deciding the spectra (SI Figure 25). The scattering by the individual droplets can be determined by Mie theory, which can be seen in the reflectivity from the dimer structure when the electric field component is perpendicular to the axis of the gap (SI Figure 26). It results effectively from the dipole excitations of the two spheres in the dimers and has a dominant contribution in the UV region (SI Figure 27). The component of spectra occurring due to the scattering should be experimentally seen as spectral features unresponsive to mechanical deformation since it is a single-particle scattering phenomenon. As expected from the Mie theory, the reflectivity peaks occurring due to scattering occurs in the UV region, as can be inferred experimentally by the unchanging features of the spectrum in the UV region (SI Figure 28). Further, the reflectivity trend is confirmed to have non-changing features in the UV region while stretching the samples fabricated with other deposition parameters and by FDTD simulations (SI Figure 29).

## Applications

With such flexibility of sample processing, various dynamically tuneable photonic devices based on plasmonic properties can be engineered and processed. As shown above, any mechanical deformation resulting in the change of inter-droplet gap will result in the corresponding color change, thus making the sample responsive to local and global strain variation. The minimum strain detected via chromaticity change is subject to the lowest stress required to effect a change in the interparticle gap. Mechano-responsiveness of color enables the qualitative gauging of strain and stress and quantitative measurements when images are coupled to the hue channel of color space or the chromaticity coordinates (SI section 4.1, SI Figure 30, 31). Thus, it removes the separation between the signal acquisition system and the corresponding display system and allows for real-time visualization of relevant signals. Our experiments demonstrate that the sensors we created are suitable for detecting both local and global strains (SI video 4), as well as monitoring healthcare (SI video 5) and detecting body motion (SI video 6). These results indicate that our sensors have the potential to be used in prosthetics and soft robotic systems.

Figure 6a depicts the mechano-responsiveness of the curvature. When the sample was bent in the lateral direction, the color changed from pink to yellow (tensile and positive curvature) and pink to blue (compressive and negative curvature). The local change in the curvature of the surface on which Ga has deposited results in color variation (SI Figure 32), as depicted in the hue plot (Figure 6a, bottom). The hue circle shown in Figure 6a indicates that a significant fraction of the hue is encompassed by dynamic tuning of the curvature of the sample.

Figure 6b exhibits the sensitivity of the sample to the point stress (SI video 7). Here, the sample is subjected to point force by a tweezer, thus resulting in a visibly distinct color change from pink to green. The significant change in the hue, as depicted in the surface plot, holds the potential to obtain the local stress varying in a region within 25 mm$^2$ or lower, determined by the source of the force, as shown in the line plot (Figure 6b).

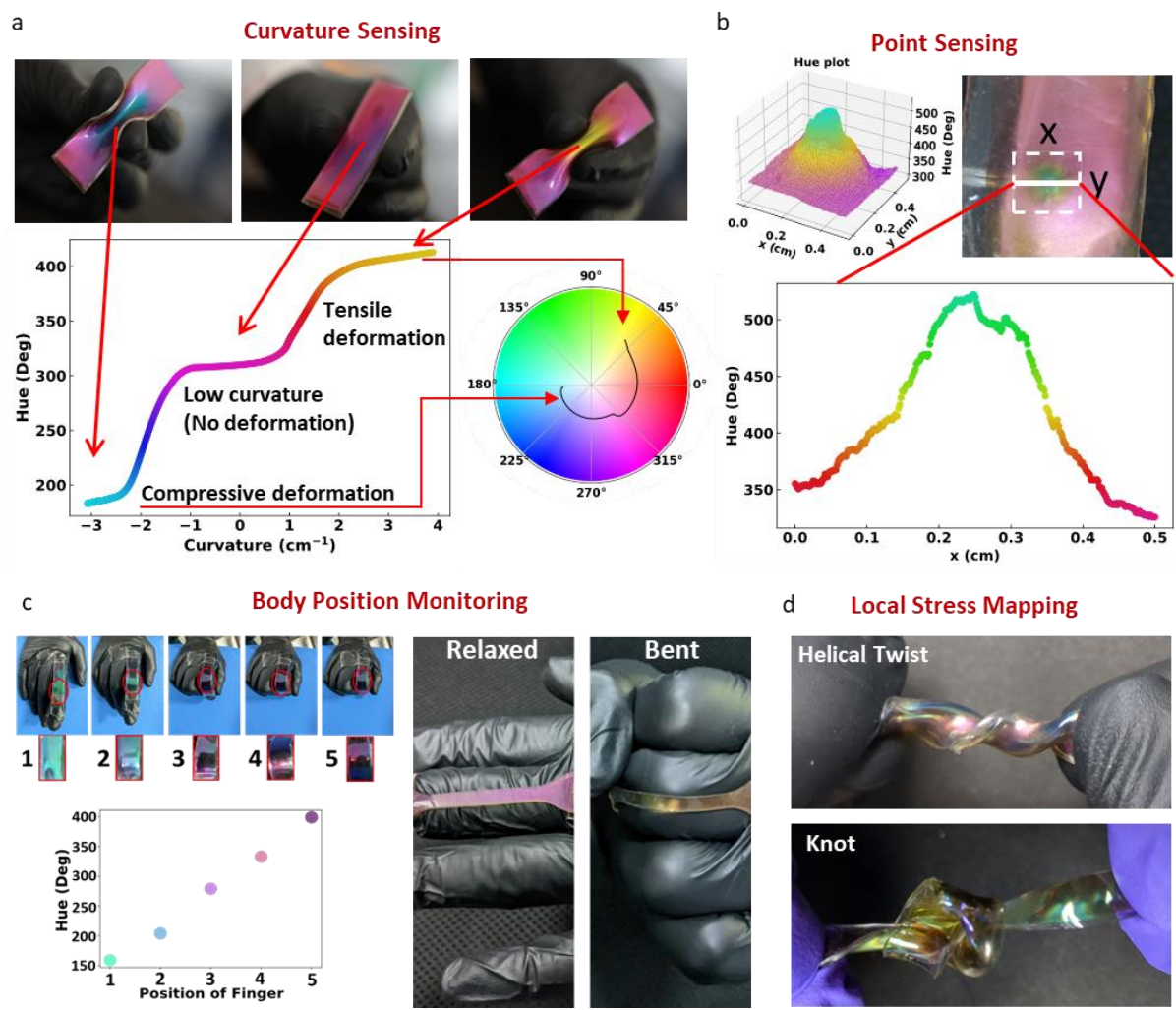

*Figure 6 | Mechano-responsive applications of Ga-deposited PDMS. a, Curvature sensing, Hue (deg) vs. Curvature of the sample. (Inset) Hue circle depicting the line along which the curvature varies. b, Color change due to point stress applied by a tweezer tip. The 2D surface plot illustrates Hue's spatial variation in the sample region marked by the dashed white boundary. (bottom) Hue vs. distance plot along the line drawn by solid white on the sample. c, Chromatic responsiveness applied in detecting the bending of fingers. d, Local variation in color of the sample due to a helical twist (top), and a knot(bottom), indicating local strain map.*

While the wide range of structural coloration can be used for static reflective displays like logos and signboards, the mechano-responsive property can be employed for sensing applications. Mechano-responsiveness in soft substrates is desirable while making human wearable interfaces for devices such as a body movement sensor (Figure 6c) (SI video 6). Here, the curvature change of the sample results in a change of color, thus, resulting in different chromaticity for different bending angles of the finger, proving it to be useful in making maneuvering systems that employ bending. In addition, the color of the flexible device can be optically analyzed to detect local strain variations, allowing for real-time stress mapping on a mechanically strained sample, like a helical twist (SI video 8) or a knot (SI video 9) (Figure 6d).

**Conclusion**

By exploiting the capillary interaction of liquid Gallium and the oligomers interspersed in the PDMS crosslinked matrix, we have demonstrated a single-step, simple, and scalable fabrication technique to realize chromogenic structures. Our proposed method generates Gallium nanoparticles through a simple physical vapor deposition process. The uniqueness of this method is that it forms non-coalescent droplets separated by a few nanometers, creating a high field intensity that excites Gap plasmon modes. The structure's color is rendered via wavelength-dependent scattering of the incident light and the Gap plasmonic modes.

Not only does the fabrication process incur a lower fabrication cost due to a single process step, but it also ensures minimal waste by-products, thus making it a sustainable method of large-scale fabrication of structural color. The span of chromaticity coordinates obtained by the self-assembly method can be used for potentially numerous applications using reflective displays, aesthetic crafts, decorations, and long-lasting smart windows.

Our experiments and simulations establish that mechano-responsiveness results from tuning the gap between the droplets by mechanical deformation of the PDMS substrate. The chromogenic structure created by the proposed method has a mechano-responsive property that could lead to affordable visual sensors and advancements in healthcare technologies, soft robotics, and smart manoeuvring systems. The reliability and repeatability of the structure hold the promise of sustainability, longevity, and durability in the respective device applications.

## Acknowledgments

The authors acknowledge the Prime Ministers Research Fellowship (TF/PMRF-21-1343), MHRD, SERB grant (SP/SERB-22-0021.05), DST India, and NPDF grant (PDF/2021/000865), DST India for funding the project. The authors thank Prof. Dr. Eric Dufresne and Dr Robert Style, ETH Zurich for the discussions, which helped understand the fluid mechanics behind the formation of Ga nanodroplets. They also thank the Micro and Nano Characterization Facility (MNCF), CENSE and Advanced Facility for Microscopy and Microanalysis (AFMM), IISc, Bangalore. They appreciate the support of the Multiscale Transport and Energy Research Laboratory, IISc Bangalore, for the measurement of interfacial properties of solids and liquids, the Biomechanics Laboratory, IISc Bangalore, for the preparation of UTM samples and the Laboratory for Advanced Manufacturing & Finishing Processes, IISc Bangalore for the fabrication of the experimental setup.

## Author Contributions

TDG proposed the research direction, supervised the project, and participated in selecting the material and optical modeling. RRS was involved in the fabrication, characterization, and mathematical modeling of the nanostructure, optical experiments, their corresponding simulations, programming for data analysis, and hardware interfaces. ASR was involved in developing the mathematical model and experiments for fabrication, characterization, and determining the substrates' mechanical properties and measuring interfacial properties. SB participated in fabricating samples, developing reflective displays, designing and manufacturing mechanical stretching equipment, and programming it with Arduino UNO microcontroller. MV was involved in fabricating samples and developing reflective displays and sensors for body part monitoring. RRS, ASR, and TDG wrote the manuscript. All authors gave final consent to the manuscript.

## Methods

### Fabrication of Samples

*Preparation of PDMS and deposition of Ga*

Polydimethylsiloxane (PDMS) soft substrates are prepared by mixing the liquid PDMS base and crosslinking agent in various ratios (Dow Corning, Sylgard 184). The following notation PDMS XX represents 1 part of the curing agent is added to XX parts of liquid PDMS base in weight ratio. Substrate softness increases with an increase in the proportion of liquid PDMS base. To fabricate structural colors, PDMS 5, PDMS 10, PDMS 15, and PDMS 20 are chosen as soft substrates. The mixture is stirred thoroughly and desiccated to remove the bubbles. The solution is poured onto a Polystyrene (PS) petri-dish and cured at 80°C in an oven for 2 hours to attain a cured soft substrate, which inherently has some uncrosslinked liquid PDMS chains. Liquid Gallium (Thermo Scientific Chemicals, 99.999% metal basis, packaged in polyethylene bottle) is then thermally evaporated (HHV thermal evaporator) and deposited onto these substrates to form nanodroplets. The thickness and temperature of the substrates are monitored via inbuilt thickness and temperature monitoring sensors, respectively.

*Toluene Treatment of PDMS substrates*

To examine the role of uncrosslinked liquid PDMS chains, cured PDMS substrates are stirred in a toluene bath for a specific duration of time. This process dissolves the uncrosslinked liquid PDMS chains since toluene acts as a suitable solvent. During this period, toluene is changed in an interval of 24 hours. Subsequently, the PDMS substrate is immersed in ethanol for 12 hours and kept in a vacuum oven at 70°C for 12 hours to remove the solvent in the sample.

**Simulation of Gallium Nanostructures on PDMS**

Finite difference time domain algorithm in the commercially available software Lumerical is used for simulations. The size distribution of Gallium nanodroplets is determined from the SEM images. The radii of the Ga nanodroplets in a given layer are sampled from the size distribution, while their location is sampled from the uniform distribution spanning the top (xy surface, +ve x normal) of the PDMS substrate. The PDMS substrate is a homogenous material with a refractive index of 1.4. The optical constants for Ga are imported as the real and imaginary parts of the refractive index. The FDTD region is chosen as 1μm x 1μm, with periodic boundary conditions along the x and y directions. The area chosen is large enough to encompass randomly distributed Ga droplets so that periodicity effects do not come into play significantly.

The Total Field Scattered Field (TFSF) source is used for the dimer simulation with an inter-particle gap. The frequency-domain field and power monitor are put behind the injection plane of the TFSF source to obtain the back-scattered power for understanding the spectral trends. For the FDTD boundary condition perfectly matched layer (PML) is used on all sides. To obtain intensity plots, a frequency-domain field profile monitor is placed parallel to the injection plane, containing the centers of both spheres.

**Mechanical stretching of Gallium Nanostructures on PDMS**

Gallium-deposited PDMS substrate is cut into a 5cm x 1cm rectangular shape. Both the sample ends are clamped onto stretching equipment operated through a Stepper motor controlled by Arduino. It is programmed for linear stretching as well as for applying a periodic strain.

**Material Characterization**

*Scanning Electron Microscopy (SEM):* The scanning electron microscope images are taken with Zeiss-Gemini SEM, ULTRA 500 model in the inlens and secondary electron mode. The samples are sputtered with 10nm gold to form a conducting layer for electronic charge dissipation.

*Focussed Ion Beam (FIB):* Thermo Fisher Scientific Scios Dual Beam FIB-SEM system is used to make thin lamella, which is placed on a copper grid for cross-section imaging in secondary electron mode.

*Scanning Tunnelling Electron Microscopy (STEM):* ThermoFisher® Titan™ Themis Scanning Tunnelling Electron Microscope is used to obtain high-angle annular dark-field scanning transmission electron micrographs (HAADF STEM) using a field emission gun with an extractor voltage of 300kV. The elemental composition of Ga embedded in PDMS was obtained by Energy Dispersive X-ray (EDAX) elemental mapping using a super X detector in STEM nanoprobe mode in TITAN.

**Optical characterization**

The Ocean Optics FLAME series UV-Visible spectrophotometer is used to obtain the reflectivity at normal incidence for each sample. The package python-Seabreeze is used to access the spectrophotometer in a python program. The reflectivity thus obtained is used for real-time computation of the sample's CIE x and y coordinates and hue. The packages numpy and matplotlib are used for array manipulation and visualisation (graphs, animations) respectively.

# Supplementary Information

Scalable fabrication of gap-plasmon-based dynamic and chromogenic nanostructures by capillary-interaction driven self-assembly of liquid-metal

Renu Raman Sahu, Alwar Samy Ramasamy, Santosh Bhonsle, Mark Vailshery, Tapajyoti Das Gupta*

Laboratory of Advanced Nanostructures for Photonics and Electronics, Department of Instrumentation and Applied Physics, Indian Institute of Science, C. V. Raman Road, Bengaluru-560012, India

E-mail: tapajyoti@iisc.ac.in

TABLE OF CONTENTS

# 1. Methods

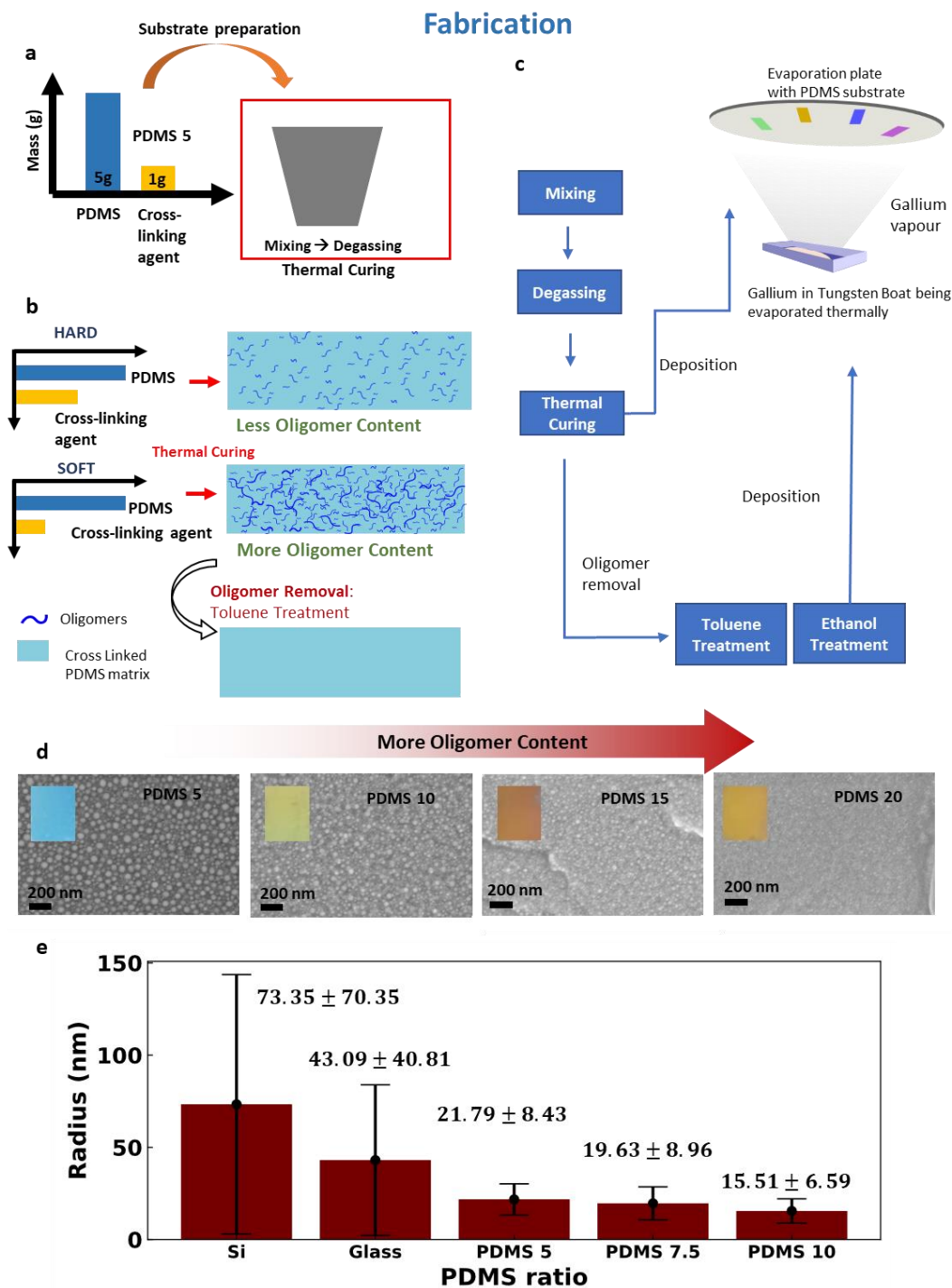

*Figure 7: a, Schematic of Fabrication of different PDMS ratios. b, Schematic diagram of the PDMS substrates fabricated with different ratios of PDMS base and curing agent, and the effect of Toluene treatment on the PDMS substrate. c, Thermal evaporation of Gallium. d, The top view SEM images of PDMS ratios 5, 10, 15, 20 (from left to right). The optical images of Ga deposited on the substrates are shown in the inset. e, Average and standard deviation of radii of Ga nanodroplets formed on the substrates Si, Glass, PDMS 5, 7.5 and 10 after thermal evaporation.*

## 1.1 Preparation of PDMS

Polydimethylsiloxane (PDMS) soft substrates are prepared by mixing the liquid PDMS base and curing agent in various ratios (Dow Corning, Sylgard 184). The following notation PDMS XX represents 1 part of curing agent is added to XX parts of liquid PDMS base in weight ratio. Substrate softness increases with an increase in the proportion of liquid PDMS base. To fabricate the structural color from Ga, PDMS 5, PDMS 10, PDMS 15, and PDMS 20 are chosen as substrates. The mixture is stirred thoroughly and desiccated to remove the bubbles. The solution is poured onto a Polystyrene (PS) petri-dish and cured at 80°C in an oven for 2 hours to attain a cured soft substrate, which inherently has some uncrosslinked liquid PDMS chains. Liquid Gallium is then thermally evaporated (HHV thermal evaporator) and deposited onto these substrates to form nanodroplets. The thickness and temperature of the substrates are monitored via inbuilt thickness and temperature monitoring sensors, respectively.

## 1.2 Toluene Treatment of PDMS substrates

To examine the role of uncrosslinked liquid PDMS chains, cured PDMS substrates are stirred in a toluene bath for a specific duration of time. This process dissolves the uncrosslinked liquid PDMS chains since toluene acts as a suitable solvent. During this period, toluene is changed in an interval of 24 hours. Subsequently, the PDMS substrate is immersed in ethanol for 12 hours and kept in a vacuum oven at 70°C for 12 hours to remove the solvent in the sample.

## 1.3 Elastic Modulus of PDMS substrates: with and without Oligomers

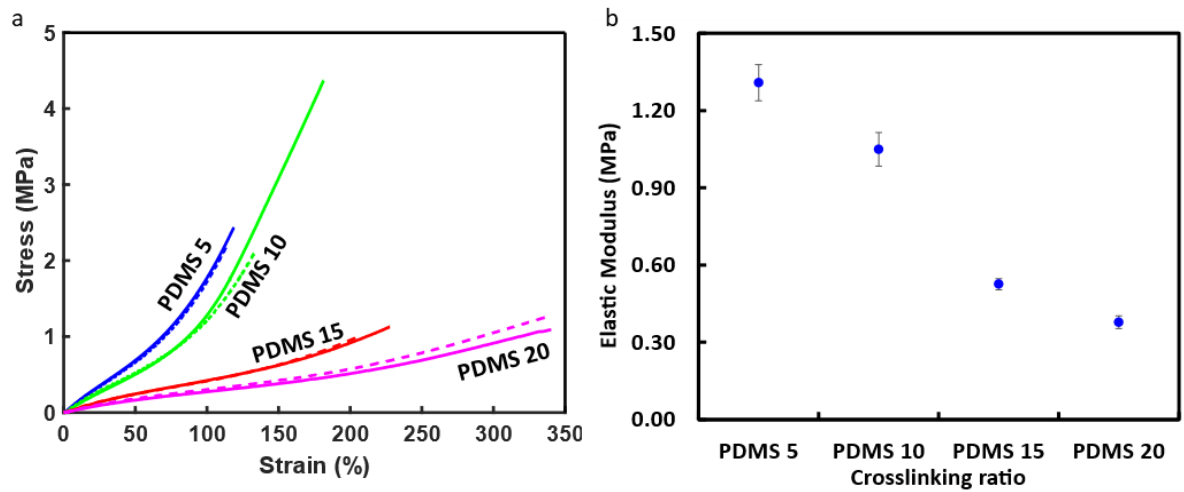

*Figure 8: a, Stress-Strain plots of different PDMS samples with oligomers (solid line) and without oligomers (dotted line) subjected to a tensile strain in the Universal Testing Machine (UTM). b, Elastic modulus of different PDMS crosslinked substrates measured from the stress-strain curve with upto 40% strain, where the stress and strain are linear.*

## 1.4 Effect of liquid oligomers of PDMS on the Gallium nanostructures

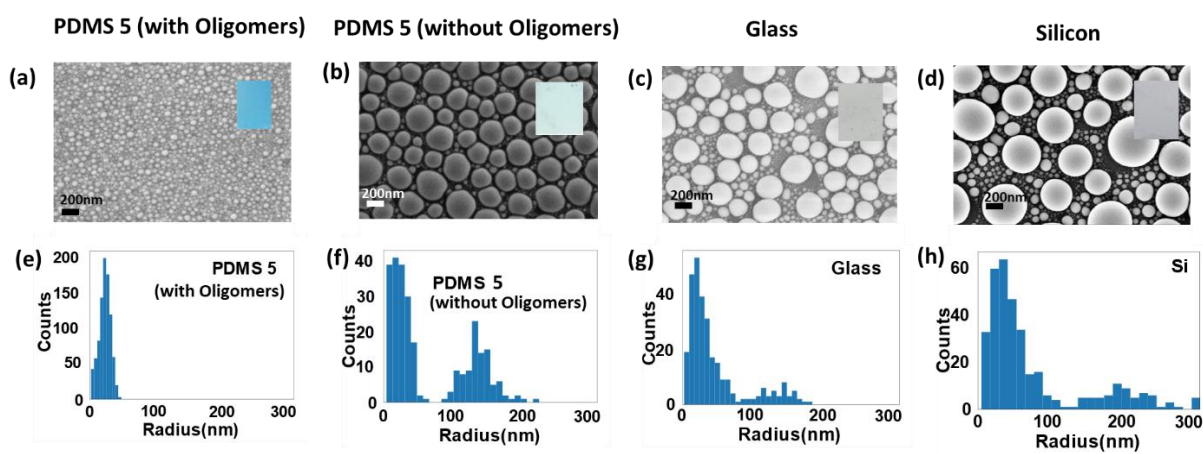

Figure 9: Top view SEM images when Ga is deposited onto (a) PDMS (with oligomers), (b) PDMS 5, (without oligomers), (c) Glass and (d) Silicon. The size distribution of Ga nanodroplets formed on the substrates (e) PDMS 5 (with oligomers), (f) PDMS 5, (without oligomers), (g) Glass and (h) Silicon.

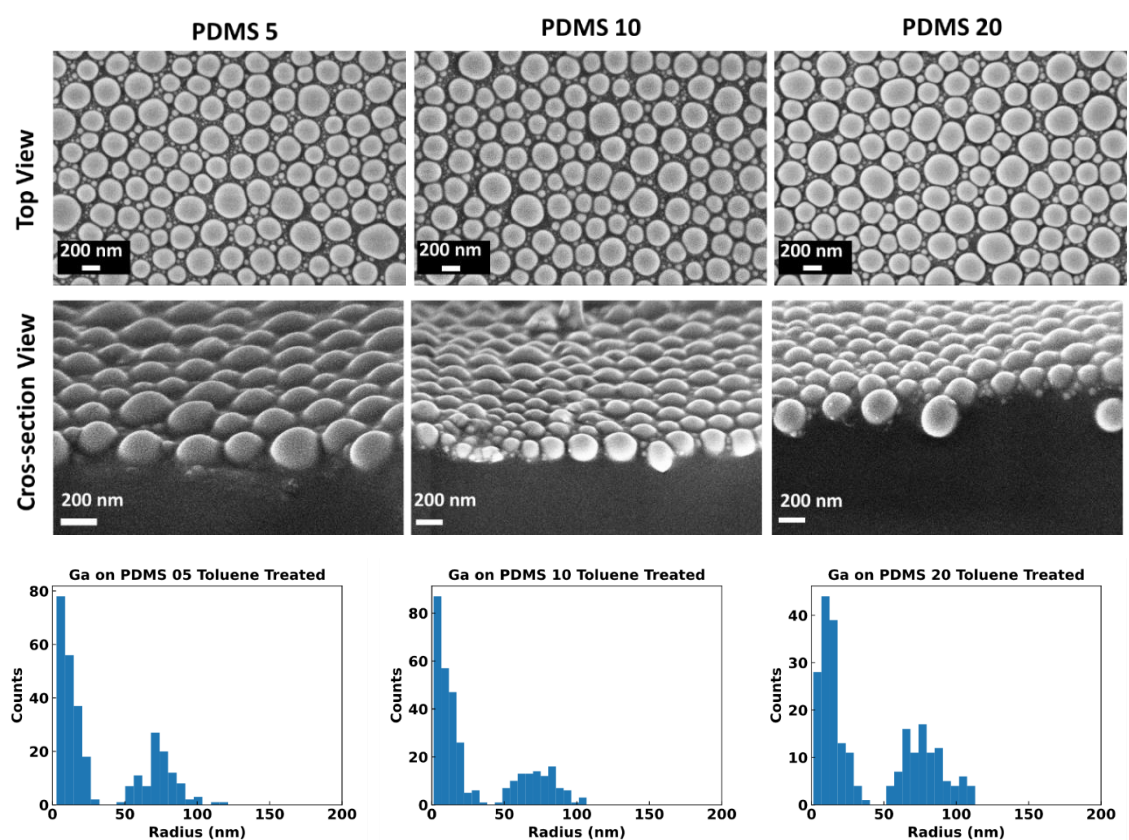

Figure 10: Top-view (top panel) and cross-sectional (middle panel) SEM images of Ga nanostructures on PDMS 5, 10, and 20 after removal of oligomers by toluene treatment and their corresponding size distribution is shown in the bottom panel.

The size distribution of Ga nanoparticles on PDMS substrates without oligomers is similar, characterized by large droplets of radii varying between 50 and 100 nm, and small droplets with radii less than 50 nm. Although Young's modulus of the substrates varies (SI Figure 2), the absence of liquid oligomer content eliminates the difference in the size distribution of the Ga nanodroplets. This establishes that the fluidic interaction between the liquid oligomers and liquid Ga droplet results in a layered nanodroplets structure, thus producing colors.

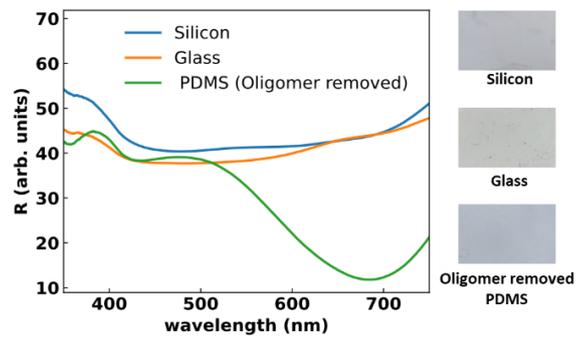

*Figure 11: Reflectivity spectra of Ga deposited Si, Glass and Toluene treated PDMS. (Right column) Optical images of Ga deposited simultaneously on Si, glass and PDMS substrates without oligomers (Toluene treated).*

Thus, we conclude that removing oligomers eliminates the distinction between different PDMS ratios. Therefore, oligomers play a crucially significant role in the determination of nanostructures.

# 2. Mathematical modeling of Ga nanostructure formation on PDMS: Substrate, Growth and Engulfing Equations

In this section, we hypothesize the formation mechanism of Ga nanospheres and their embedding into the PDMS. Also, we propose a set of differential equations based on a few assumptions to model the experimentally observed size distribution trends of Ga nanodroplets.

## 2.1 Nucleation and formation of droplets of critical radii

Thermal evaporation of Gallium onto the substrate results in Ga vapors near the PDMS substrate, which condenses to form nanospheres. The surface tension of Ga renders the droplets to be spheres. The formation of a Liquid Ga-air interface costs energy proportional to the surface area of the sphere, whereas condensation into liquid droplets releases its vaporisation energy proportionate to its volume. Therefore there would be a critical radius which would be the minimum radius of Ga nanodroplets that form upon condensation. Once the Ga nanodroplets form, they grow due to further condensation of Ga vapor to the liquid droplet of a few nanometers radius before it comes into contact with the PDMS substrate.

## 2.2 Initial deformation of the substrate due to Laplace pressure

When the nanodroplets come in contact with the substrate, the initial speed it and its Laplace pressure will result in the substrate's initial deformation, leading to a small immersion of the nanodroplet into PDMS. Once infinitesimal immersion happens, the contact line of Ga droplet and PDMS occurs, and the following sequence of events occur.

Young's law governs the wetting of droplets on rigid substrates and Neumann's triangle on liquid substrates, which is infinitely soft[1]. In between the rigid and liquid substrates called the soft solids, the contact angle for larger droplets follows Young's law, whereas, with the smaller droplets, the contact angle is obtained by the balance of Neumann's triangle. In soft solids, the balance between capillarity and elastic forces is governed by the elastocapillary length scale $l_{el} = \frac{\gamma_{Ga}}{E}$, where $\gamma_{Ga}$ is the surface tension of the gallium droplet, and $E$ is Young's modulus. For the magnitude of radius $R > l_{el}$, ridge formation at the contact line is small, and macroscopically droplets follow Young's law like a rigid substrate. Whereas for small droplets $R < l_{el}$, high Laplace pressure inside the droplet can deform and form a dimple on the substrate. In such cases, the droplet profile takes the shape of a liquid lens which is given by the balance of $\gamma_{Ga}$, $\Upsilon_{PDMS}$ and $\Upsilon_{PDMS-Ga}$. The elastocapillary length for a Ga droplet on PDMS 5 is 496. During typical evaporation, the radius of the Ga droplet is observed to be around 25 nm. Hence the droplet will resemble the case of a liquid lens on PDMS substrates.

## 2.3 Assumption of spherical geometry of partially immersed Gallium nanodroplets

The Ga droplet on PDMS can be considered to be placed equivalently on a liquid substrate from the elastocapillary length scale analysis (Sec 2.2). As a result, the shape of a Ga droplet on a PDMS substrate is governed by the balance of interfacial stresses (surface tension forces) at the three-phase contact line, which is well characterized by Neumann's triangle[2]. The balance of forces in horizontal and vertical directions gives,

$$\gamma_{Ga} \cos \alpha + \gamma_{Ga-PDMS} \cos \beta - \gamma_{PDMS} \cos \sigma = 0 \qquad (4)$$

$$\gamma_{Ga} \sin \alpha - \gamma_{Ga-PDMS} \sin \beta + \gamma_{PDMS} \sin \sigma = 0 \qquad (5)$$

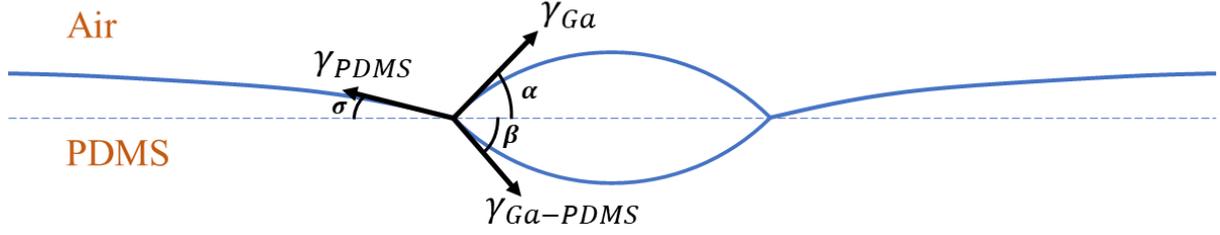

*Figure 12: Schematic of Ga droplet on PDMS substrate for very high elastocapillary length compared to the droplet radius.*

From the cosine law of the triangle, relation between the angle made by the different liquid with horizontal and surface energies can be expressed as,

$$\alpha + \beta = \cos^{-1} \frac{\gamma_{PDMS}^2 - \gamma_{Ga-PDMS}^2 - \gamma_{Ga}^2}{2\gamma_{Ga}\gamma_{Ga-PDMS}} \tag{6}$$

$$\alpha + \sigma = \pi - \cos^{-1} \frac{\gamma_{Ga-PDMS}^2 - \gamma_{Ga}^2 - \gamma_{PDMS}^2}{2\gamma_{Ga}\gamma_{PDMS}} \tag{7}$$

From the geometrical analysis, the volume occupied by Ga with the Gallium-air and PDMS-Gallium interfaces can be calculated as,

$$V_1 = \frac{\pi R_d^3}{3\sin^3\alpha}(1-\cos\alpha)^2(2+\cos\alpha) \tag{8}$$

$$V_2 = \frac{\pi R_d^3}{3\sin^3\beta}(1-\cos\beta)^2(2+\cos\beta) \tag{9}$$

Equating the volume in the Ga of occupied with two different arcs of radius to the volume of the initial spherical droplet $V_1 + V_2 = V$, contact radius $R_d$ can be found. With the contact radius, radius of the lens formed by the Ga droplet can be determined by,

$$R_1 = \frac{R_d}{\sin\alpha} \tag{10}$$

$$R_2 = \frac{R_d}{\sin\beta} \tag{11}$$

Since the surface energies of Ga-air and Ga-PDMS interface are close, the exposed and immersed radius of curvature is almost the same. For example, for a Ga droplet of 200 nm diameter, the exposed and immersed radii are 102.4nm and 100nm. This enables us to assume a spherical geometry of Ga nanodroplet even during immersion into the PDMS.

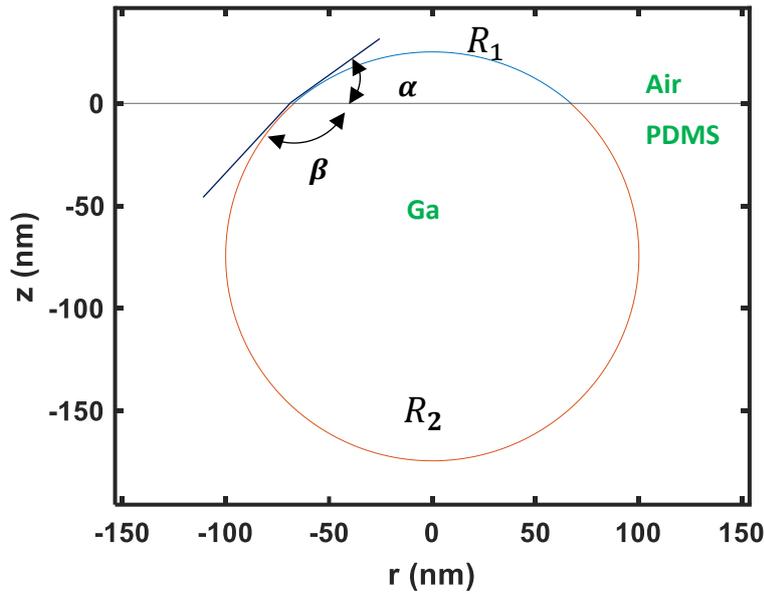

*Figure 13: The shape of the Ga droplet (200 nm diameter) on the PDMS substrate is calculated analytically. Gravity is neglected at the nanoscale, leading to equal pressure across both air and PDMS phases and hence the curve remains a flat line between both phases. From Neumann equilibrium conditions, contact angles are obtained, $\alpha = 40.8°$ and $\beta = 138°$. Using the angles and volume conservation, the radius of the spherical arcs are $R_1 = 102.4\ nm$ and $R_2 = 100\ nm$.*

**2.4 Positive spreading Parameter: Engulfing of Ga droplets layer**

The phenomena of fluid separation and ridge formation occur when materials with a positive spreading parameter (defined in the manuscript) are used to swell PDMS[3,4]. The spreading parameter determines the minimum energy configuration of three interfaces. For our case, the spreading parameter $S$ turns out to be

$$S = \gamma_{Ga-air} - \gamma_{Ga-oligomers} - \gamma_{oligomers-air} = 650 - 590 - 20 = 40\ mN/m \quad (12)$$

Here $\gamma_{Ga-air}$, $\gamma_{Ga-oligomers}$ and $\gamma_{oligomers-air}$ denote the surface tensions between Ga and air, Ga and oligomers, and oligomers and air. A positive spreading parameter of 40mN/m implies that the surface energies of Ga-air interface is 40 milliJoule per square meter more energy than that of the two interfaces Ga-oligomers and oligomers-air, combined. To minimize the surface energy of the system (for a positive spreading parameter), the oligomers will separate from the PDMS network and cloak the Ga nanodroplets. The mechanism of Gallium nanodroplets penetrating the PDMS is hypothesized here using the result for positive spreading parameter and a few assumptions made about the substrate and its interaction with Ga nanodroplets. The positive spreading coefficient ($S > 0$) of the system results in the extraction of oligomers from the cross-linked PDMS network and tends to cloak the nanodroplet. During engulfing, the nanodroplet pulls the oligomers to cloak it up. The surface undulations of the oligomers are energetically unfavorable as compared to a planar surface. Therefore, the surface tends to be planar, thus extracting more oligomers out of the cross-linked PDMS network. Moreover, the initial velocity of the nanodroplet into the PDMS contributes to their immersion.

**2.5 Hypothesis of engulfing mechanism and Substrate-Growth-Engulfing (SGE) equations**

Engulfing of the Ga nanodroplet is an interplay of the following three events co-occurring.

a) Separation of Oligomers from the PDMS network

Let $f$ be the volume of liquid oligomers present in a given substrate volume in units of nm³. The rate at which $f$ changes when a Ga droplet-oligomer interaction takes place dictates the phenomena of oligomer separation and will be described by the expression $\frac{df}{dt}$.

b) Growth of Ga droplet

The rate of increase of radius $R$ (in units of nm) Ga nanosphere describes its growth and is depicted by the expression $\frac{dR}{dt}$.

c) Engulfing of Ga nanodroplet

We define the engulfing angle θ (radian) as the polar angle of the contact line Ga-air-PDMS interface if one assumes the center of the sphere to be the origin and positive z-axis as the normal of undeformed PDMS-air interface from PDMS to air (SI Figure. 8). The rate $\frac{d\theta}{dt}$ describes the engulfing of Ga-droplet.

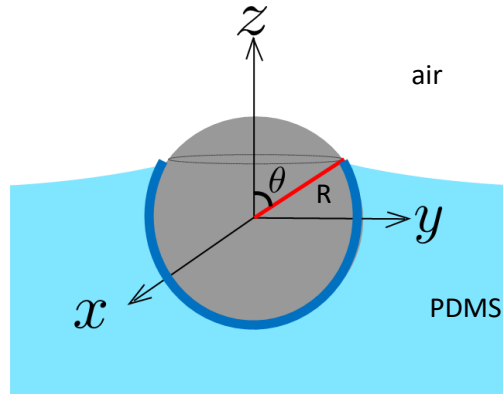

*Figure 14: A partially immersed and engulfed Ga nanodroplet with an engulfing angle θ. Note that the exposed surface area is $S = 2\pi R^2(1 - \cos\theta)$ and that immersed is $2\pi R^2(1 + \cos\theta)$. When $\theta = 0$, it is the case of complete immersion. A completely immersed nanodroplet cannot grow, and hence the rate of change of radius vanishes. No,te that $S = 0$ at $\theta = 0$, indicating complete immersion. At $\theta = \pi, S = 4\pi R^2$, the entire surface is exposed and hence the rate of increase of radius is maximum.*

In the following sections, we present the proposed model of these events in terms of differential equations. This set of differential equations describes the substrate-droplet interaction of a single nanodroplet.

**2.6 Substrate Equation**

Oligomers possess fluidic properties and are embedded homogeneously in the PDMS network. When a droplet touches the PDMS, the Ga-air-PDMS interface is formed and oligomers tend to diffuse from the bulk PDMS to the droplet interface and engulf it. This requires migration of the oligomers to the interface. Note that, at a given instant $f$ is the total volume of oligomers present in the PDMS network available for migration towards interface and contribute to droplet-engulfing. The unit of $f$ is that of volume (nm³). The volume of oligomers used to engulf the nanodroplet will no longer be available in the PDMS network. We assume that a thin layer of thickness $w$ nm of oligomers will engulf the nanodroplet in the immersed area uniformly. We hypothesize that the rate at which $f$ will deplete is proportional to the volume of oligomers left in the PDMS network after an engulfing of θ is,

$$\frac{df}{dt} \propto f - 2\pi R^2 w(1 + \cos\theta)$$

The migration of oligomers toward the interface occurs due to the contact force and we assume it to be proportional to $2\pi R \sin\theta$,

$$\frac{df}{dt} \propto 2\pi R \sin\theta$$

With $\kappa_s$ as the proportionality constant, we obtain

$$\frac{df}{dt} = -\kappa_s\big(f - 2\pi R^2 w(1 + \cos\theta)\big)(2\pi R \sin\theta) \quad (13)$$

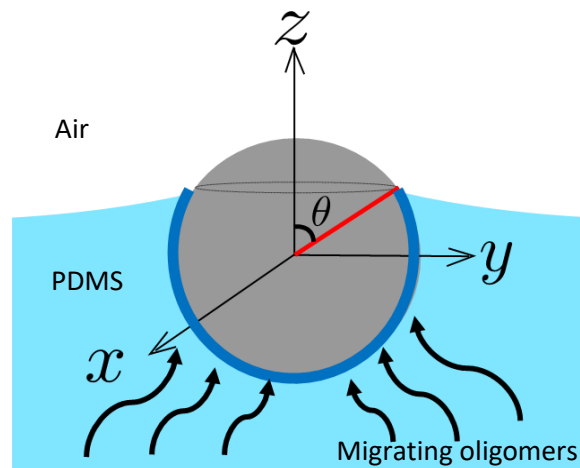

*Figure 15: Schematic of oligomers migrating from the PDMS network to the interface of PDMS and Gallium. The Ga nanodroplet will be cloaked by a thin liquid oligomer layer.*

Here $\kappa_s$ is known as substrate constant and describes the ease of migration of oligomers through the PDMS network. It is a dimensional constant with units $nm^{-1}s^{-1}$. It does not vary during the process of engulfing because the cross-linked network is assumed not to change during engulfing. The negative sign on the right-hand side signifies the reduction of oligomers as they are used up for engulfing.

Intuition dictates that higher PDMS ratios (those with a lower proportion of curing agent) have a more volumetric fraction of oligomers in a given substrate volume and hence will be characterized by higher values of $f$.

### 2.7 Growth equation

The conversion from vapor to liquid occurring at the exposed surface area results in a change in the volume of the nanodroplet. Let ρ be the density of liquid Gallium (5.9 g/cc), and $J$ the flux of liquid Ga entering the droplet through condensation. We can therefore write,

$$\rho \frac{d}{dt}\left(\frac{4}{3}\pi R^3\right) = 2\pi R^2 (1 - \cos\theta) J \qquad (14)$$

or,
$$\frac{dR}{dt} = \frac{J}{2\rho}(1 - \cos\theta) \qquad (15)$$

or,
$$\frac{dR}{dt} = \kappa_\rho (1 - \cos\theta) \qquad (16)$$

Here $\kappa_\rho = \frac{J}{2\rho}$ is the growth constant with unit $nm\ s^{-1}$.

The growth constant is determined by the flux of Ga atoms condensing into the droplet, which depends on temperature and the density of Ga atoms in the vapor phase. One can experimentally control the number density of Ga atoms in vapor phase by manipulating the deposition rate during thermal evaporation. More is the number density of Ga-atoms in the vapor phase, the faster the condensation into the liquid droplet phase.

Temperature plays a crucial role in determining the vapor-liquid equilibrium of Ga. A higher temperature favors the vapor phase, and hence the rate of condensation decreases, which causes a reduction in $J$ and a lowering of the growth constant $\kappa_\rho$.

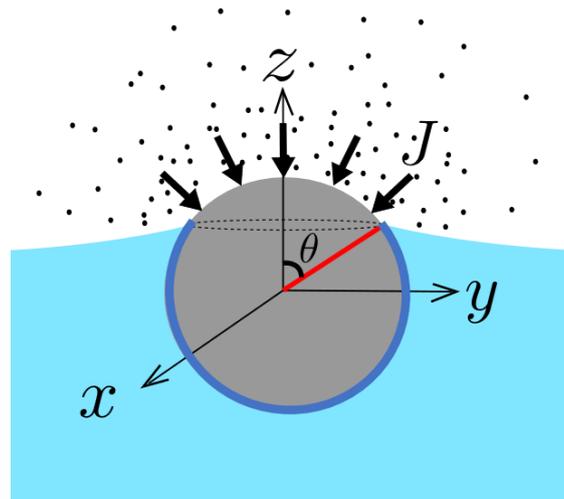

*Figure 16: Schematic of the Ga vapor liquefying at the interface of Ga vapor and nanodroplet. The flux of Ga atoms condensing into the droplet is directly proportional to the number density around the exposed surface of the Ga droplet. With the increase in temperature, the probability of Ga atoms being in a vapor state increases, thus reducing the flux of condensing Ga atoms.*

**2.8 Engulfing Equation**

The initial immersion due to the Laplace pressure of the Ga droplet renders the initial value of $\theta$ to be $\pi - h$, where $h \ll 1$. As the engulfing occurs, its value decreases towards $\theta = 0$, which is the case of complete engulfing.

We hypothesize that the rate at which engulfing occurs is proportional to the amount of oligomers available for migrating towards the interface and contributing to engulfing. Thus, we can write

$$\frac{d\theta}{dt} = -\kappa_e \left( f - 2\pi R^2 w (1 + \cos \theta) \right) \quad (17)$$

where $\kappa_e$ is the engulfing constant with unit $nm^{-3}s^{-1}$.

### 2.9 Iteration Procedure to obtain the number of immersed layers.

The substrate-growth-engulfing (SGE) equations describe the substrate droplet equation for a single droplet. The boundary conditions for the equations are defined by θ, from infinitesimal engulfing ($\theta = \pi - h$) to complete engulfing ($\theta = 0$). During this process, $f$ decreases and the radius $R$ of the nanodroplet increases. There are ample oligomers for engulfing the initial layers (the deepest layer observed in the SEM image). The rate of engulfing would be faster for the droplets in these layers; hence the growth time is relatively less than the droplets from the top layer. Here the engulfing phenomena dominate, and complete engulfing takes place. For the subsequent layers of droplets, the amount of oligomers is depleted, thus reducing the rate of engulfing and increasing the growth time. For the topmost layer, the growth time dominates the engulfing time and partial engulfing occurs.

To model the above picture, we solve the SGE equations for a nanosphere with an arbitrary but reasonable choice of initial radius and a thickness of oligomer coating around it. The initial volume of oligomers $f$ is chosen reasonably as well. The values of SGE constants $\kappa_s$, $\kappa_e$ and $\kappa_\rho$, decide complete or partial engulfing. If complete engulfing occurs, the initial value of $f$ for the next layer will be the final value of $f$ in the case of a most recent engulfed droplet. Once the partial engulfing occurs, we stop the iteration and infer the number of iterations required to obtain the partial engulfing, which is the number of layers of Ga nanodroplets formed in the PDMS.

### 2.10 Results from SGE equations

On imposing the condition that the immersion angle is monotonically non-increasing with time, solutions of the SGE differential equations for a single droplet determine whether complete immersion will occur. The value of $R$ at $\theta = 0$ (complete engulfing) is the final radius of the droplet.

To demonstrate an example, let us consider a PDMS substrate parametrized by $f_0 = 3000 \, nm^3$ and substrate constants $\kappa_s = 0.3 nm^{-1}s^{-1}$, $\kappa_\rho = 500 nm \, s^{-1}$ and $\kappa_e = 0.1 \, nm^{-3}s^{-1}$. We choose the initial radius of the droplet $R_0 = 4nm$ and the thickness of liquid oligomers engulfing it to be $w = 1nm$. The choice of the initial infinitesimal engulfing is chosen to be $\theta = \pi - h$ with $h = 10^{-4} rad$. While solving the SGE equations numerically, we deliberately impose θ to be monotonically decreasing and non-negative. Boundary values that satisfy these conditions for θ result in physically acceptable solutions.

### 1st layer

The first iteration of the numerical solution of SGE equations results in the monotonic decrease of θ from infinitesimal engulfing ($\theta = \pi - h$) to complete engulfing ($\theta = 0$). During this time, the radius of the nanodroplet increases from 4nm to 9.9nm, and the volume of oligomers decreases from $f = 3000 nm^3$ to $f = 2691 nm^3$.

The radius of the Ga droplet for the first layer is 9.9nm. The remaining volume of oligomers, $f = 2691 nm^3$ will be used as the initial oligomer volume $f_0$ for the next iteration.

:

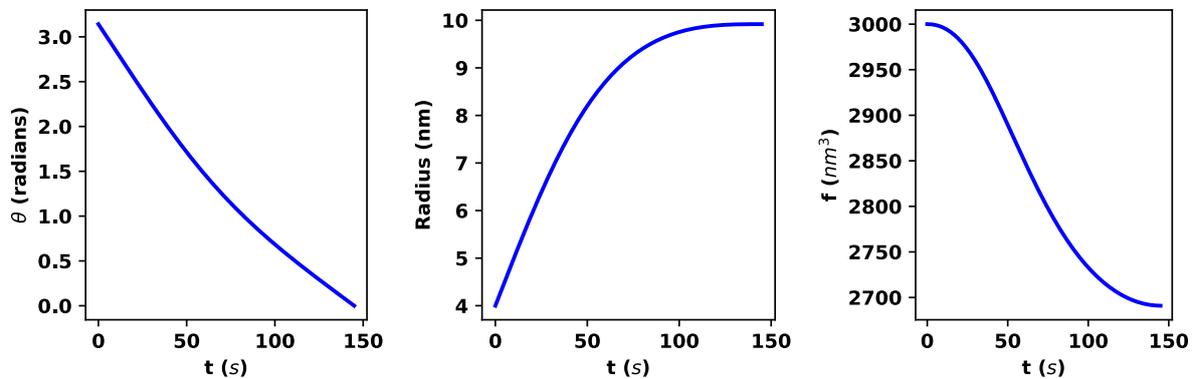

*Figure 17: Graphs of (left) engulfing angle θ vs time t. (center) Radius (R) of Ga in the first layer vs. time t. (right) Volume of oligomers f vs. time t. These are the solutions of SGE equations for the first iteration. One can infer that complete engulfing occurs here, leading to the final radius of the Ga nanodroplet being 9.9nm.*

**2nd layer**

The second iteration also results in complete engulfing, with a final Ga nanodroplet radius of 10.9 nm. Note that the radius is greater than that obtained in the previous layer. The remaining volume of oligomers $f = 2360 nm^3$ is used as the initial oligomer volume for the next iteration.

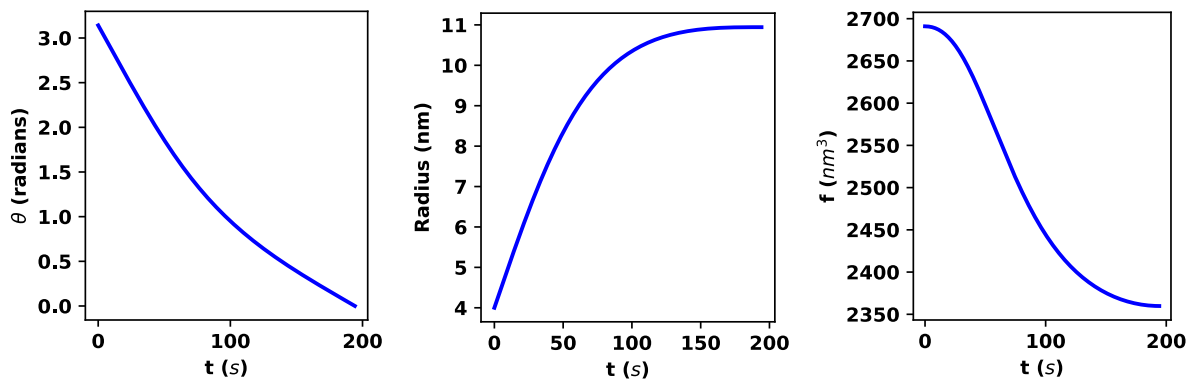

*Figure 18: Graphs of (left) engulfing angle θ vs time t. (center) Radius (R) of Ga in the first layer vs. time t. (right) Volume of oligomers f vs. time t. These are the solutions of SGE equations for the second iteration. One can infer that complete engulfing occurs here, leading to the final radius of the Ga nanodroplet being 10.9nm.*

**3rd layer**

In the third iteration, the final engulfing angle is $\theta = 0.6\ rad = 34°$, thus indicating an incomplete immersion. This would be the topmost layer of the Ga droplets deposited onto the PDMS. The droplets at this layer have a radius of 13.3 nm, which is larger than the radius of droplets in the immersed layers.

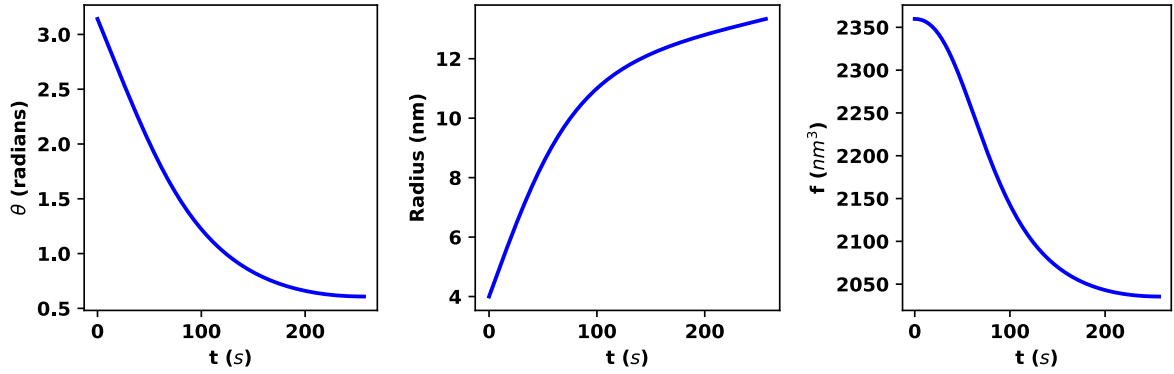

Figure 19: Graphs of (left) engulfing angle θ vs time t. (center) Radius (R) of Ga in the first layer vs. time t. (right) Volume of oligomers f vs. time t. These are the solutions of SGE equations for the third iteration. One can infer that incomplete engulfing occurs here. Hence it would be considered the topmost layer. The radius of the Ga particle formed here is 13.3 nm.

Thus, we obtain the depth profile of the Ga nanodroplet radius, as shown in the following figure.

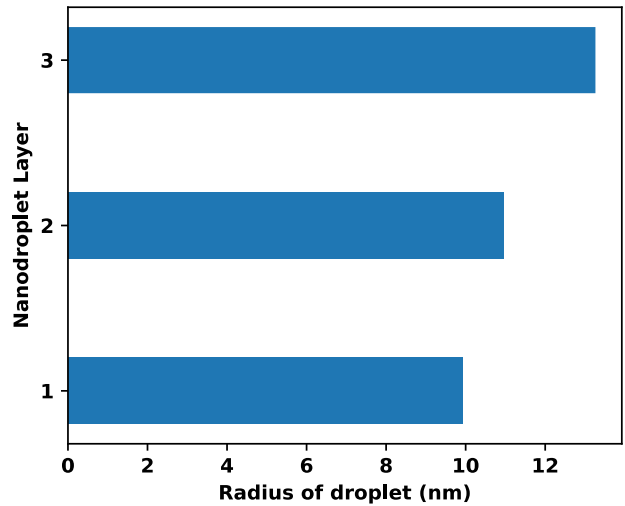

Figure 20: Histogram depicting the depth profile of Ga-radius across the cross-section. The deepest layer (1st layer) is the smallest in size. The size gradually increases with a decrease in depth. The topmost layer is of the highest radius. This trend matches with the observed cross-sectional images from SEM.

## 2.11 Comparison of the results from SGE equations with that from the experiments

As shown in the following figure, for each size distribution as observed in the cross-section images, there exist parameters of SGE equations which, when used to solve, give the matching results. The parameters used for the following three SEM images are tabulated as follows.

A note about the trend of parameters which gives the matching results for observed SEM images, are in order. The growth parameter $\kappa_\rho (= J/2\rho)$ is the same for all the PDMS substrates because the three samples were deposited simultaneously. The number density of Ga atoms in the vapor and the density of Ga are the same for all substrates during a particular deposition. The initial radius is chosen arbitrarily but is the same for all three substrates because it is formed before the droplet encounters the substrates; hence it is independent of the substrate.

The substrate constant $\kappa_s$ increases with the PDMS ratio. It measures the ease of movement of oligomers through the PDMS network. It is physically reasonable to argue that a higher PDMS ratio implies less cross-linking, making it easier for the oligomers to move through the PDMS network. More availability of oligomers means a faster engulfing, which explains the increasing trend of engulfing constant $\kappa_e$ with PDMS ratio. The thickness of liquid oligomers encapsulating the nanodroplets is inferred to increase with the PDMS ratio from SGE equations. This can be consistently accommodated by the assumption that there is ample supply of the oligomers in higher PDMS ratios and hence a more increased thickness oligomer cloaking is favorable.

| Parameters | PDMS 5 | PDMS 10 | PDMS 20 |
|---|---|---|---|
| $\kappa_s\,(nm^{-1}s^{-1})$ | 0.4 | 0.5 | 0.8 |
| $\kappa_\rho\,(nm\,s^{-1})$ | 1280 | 1280 | 1280 |
| $\kappa_e\,(nm^{-3}s^{-1})$ | 0.1 | 0.11 | 0.15 |
| $f_0\,(nm^3)$ | 4500 | 5000 | 7000 |
| $R_0\,(nm)$ | 5.0 | 5.0 | 5.0 |
| $w\,(nm)$ | 0.5 | 0.7 | 2.0 |

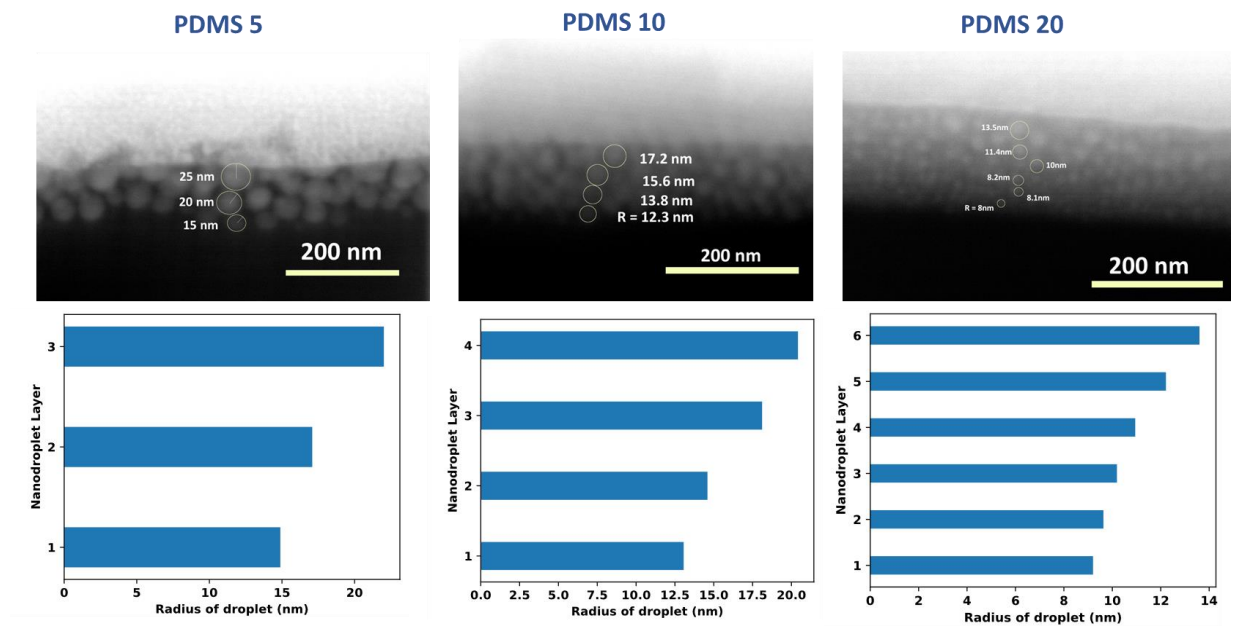

Figure 21: (Top) Table showing the parameters used for substrate-growth-engulfing (SGE) equations, matching the modeling results with the experiment's results. (Middle) The SEM images with representative circles are drawn around Ga nanodroplets to determine their radii. (Bottom) Results from SGE equations are solved using the parameters tabulated above. One can infer a fair agreement with the results from SGE equations and the observed SEM images.

## 2.12 Prediction from the SGE equation

The SGE equations have successfully explained the size distribution trends as observed experimentally. Once we have found the parameters for a particular deposition process and the PDMS substrates, we can use them and relate them with the experimental parameters to obtain a control on size distribution. For example, the rate of deposition and temperature will affect the growth constant $\kappa_\rho$. With the increase in the rate of deposition and decrease in temperature, $\kappa_\rho$ increases. Though the exact analytical expression relating $\kappa_\rho$ to these experimental parameters have not been determined yet, we can investigate the size distribution trend of varying them by changing the value of $\kappa_\rho$ in SGE equations.

For example, consider PDMS 10, with SGE parameters obtained from SEM image analysis. The following figure depicts the effect of increasing the rate of deposition, which is effectively modeled by increasing the value of $\kappa_\rho$. We expect to obtain nanodroplets of larger sizes when we increase the deposition rate or lower the temperature. In the cross-section, the number of layers reduces.

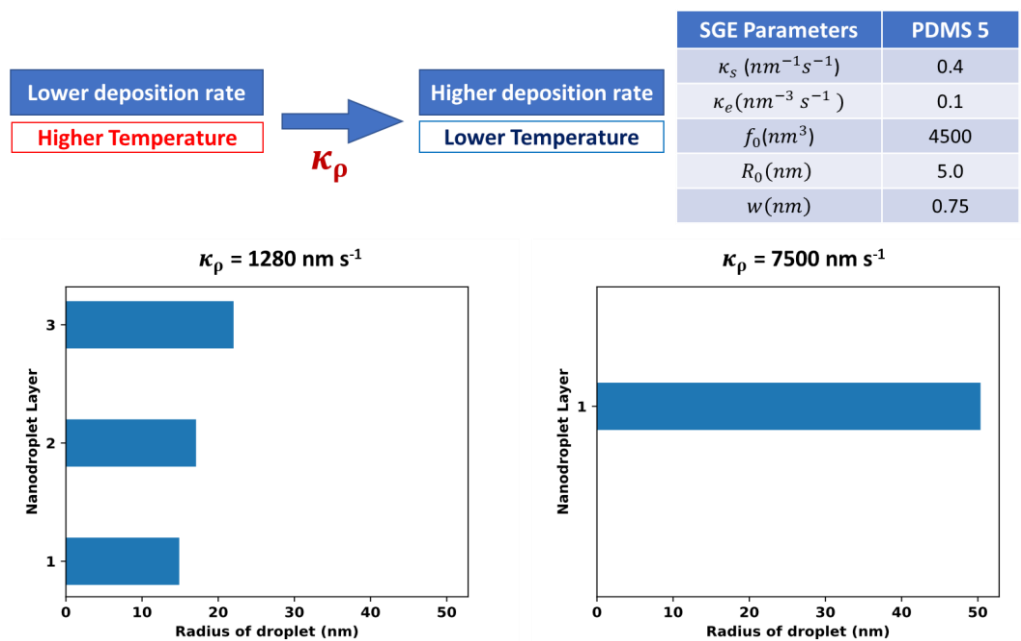

Figure 22: The predicted trend of Ga-radius deposited on PDMS with varying rates of deposition or temperature. The table shows the parameters for PDMS 10 obtained from an SEM image of the sample deposited at the rate of 1Å/s (and $\kappa_\rho = 1280\ nm/s$, the leftmost SEM image). On increasing the deposition rate or lowering the temperature, we expect to get less number of Ga layers and nanodroplets of larger sizes.

Experimentally we observed that increasing the deposition rate to 7 Å/s from 1 Å/s decreased the number of layers of Ga nanodroplets. Moreover, the cross-sectional radius of Ga nanodroplets measured from the cross-section SEM image (see Figure 15) for lower rate deposition is 20.4±3.5 nm, while the higher rate of 7 Å/s is $49.9 \pm 7.7$ nm. This observation is in tandem with the predictions from SGE equations.

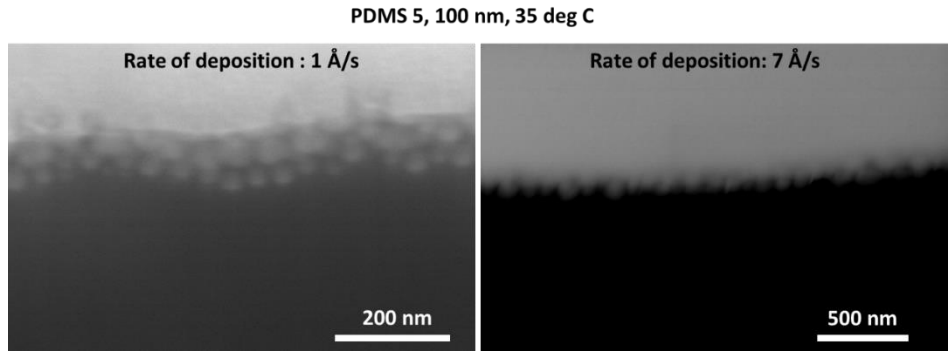

*Figure 23: Cross-sectional Image of Ga deposited on PDMS 5 at 35 deg C with deposition rate (left) 1 Å/s, and (right) 7 Å/s.*

The PDMS substrate is usually characterized by the ratio of the PDMS base to the cross-linking agent. The SGE equation provides another useful PDMS characterization utilizing SGE parameters. These parameters, as demonstrated in the previous section, describe substrate interaction with the gallium nanodroplets. It would be a novel method to describe the PDMS properties and their behavior with the Ga-deposition parameters.

## 3. Physics of Optical spectra of Ga deposited PDMS

By examining the reflectivity spectra of Ga-deposited PDMS, we can experimentally verify how light interacts with nanodroplets and gain insights into the morphology of Ga on PDMS. The results from SEM images confirm that the fluidic interaction of Gallium with the liquid oligomers of PDMS forms layers of Ga nanodroplets with varying radii. Due to encapsulation by oligomers, there will always be a dielectric Gap of refractive index 1.4 between the Ga nanodroplets. Its refractive index determines the optical properties of Ga. The real and imaginary parts of the refractive index of Ga are shown in the following figure[5].

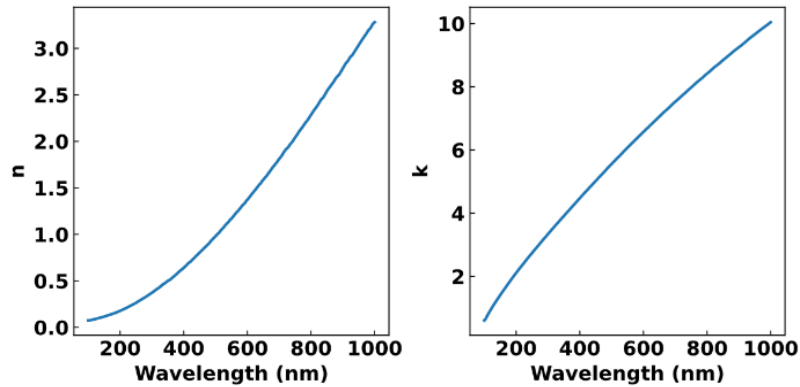

*Figure 24: The real (left) and imaginary (right) part of the refractive index of Gallium.*

Scanning electron microscopy shows that the structure of Ga nanodroplets on PDMS has the following structural characteristics:

1) The Gallium nanodroplets form several layers of nanodroplets, depending on the oligomer content of the PDMS. The more oligomer content more will be the number of layers.
2) The sizes of the Ga nanodroplets decrease with the depth of the Ga nanolayer into the PDMS.
3) An encapsulation of a thin layer of liquid oligomers separates each Ga nanodroplet.
4) For a given layer, the spatial location of Ga nanodroplets is random.

With the above constraints, we simulated some of the structures in Lumerical, a commercial FDTD software and observed a good match with the experimental reflectivity spectra.

FDTD was used with periodic boundary conditions along the x and y directions to simulate a large area of randomly distributed Ga nanostructures. The x-span and y-span were made large enough to cover a statistically significant number of Ga nanodroplets in each layer. Nevertheless, the periodicity would cause discrepancies between the experimentally obtained and the simulated reflectivity spectra. After achieving a satisfactory correlation between the experimental and simulated reflectivity spectra, the next step is comprehending the patterns observed in the reflectivity spectrum.

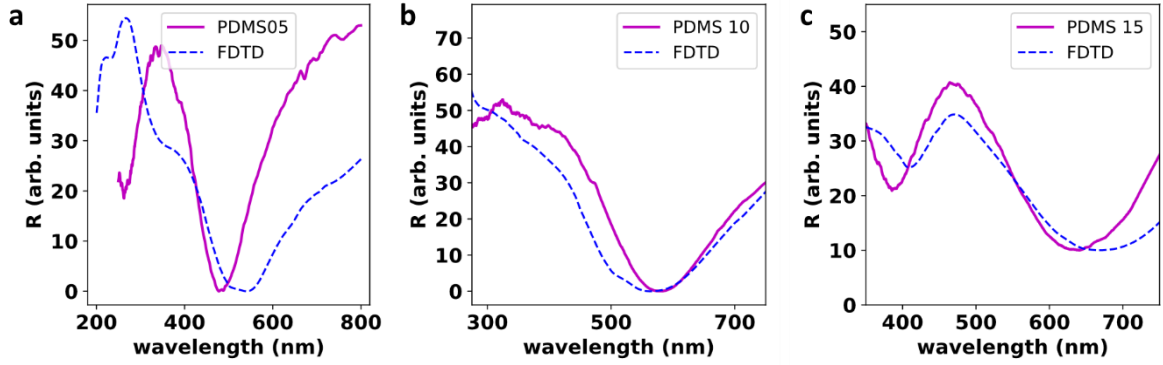

*Figure 25: Lumerical simulation of FDTD region and the experimentally obtained reflectivity spectra of (a) PDMS 5, (b) PDMS 10 and (c) PDMS 15.*

## 3.1 Variation of the spectrum with oligomer content

We observe that the spectral features of the reflectivity spectrum red-shift with an increase in the oligomer content (Figure 19). The observation is counter-intuitive given that higher oligomer content results in smaller particles; hence, one would expect a blue-shift. However, the electromagnetic fields in the inter-layer spatial region contribute to the spectrum, thus resulting in a red shift.

Although the sizes of Gallium nanodroplets decrease with the increase in the PDMS substrate's oligomer content, the spectral features consider the number and depth of the Ga nanodroplet layers. The structure with larger Ga nanodroplet layer depth results in red-shifted spectral features.

The intensity plots below depict the interaction of the incident electromagnetic field with different layers of Ga nanodroplets. At lower wavelengths, the electromagnetic field can interact with only the topmost layer of Ga spheres (see Figure 20g). In contrast, the higher wavelength fields can interact with all the layers of the structure (see Figure 20h and 20i).

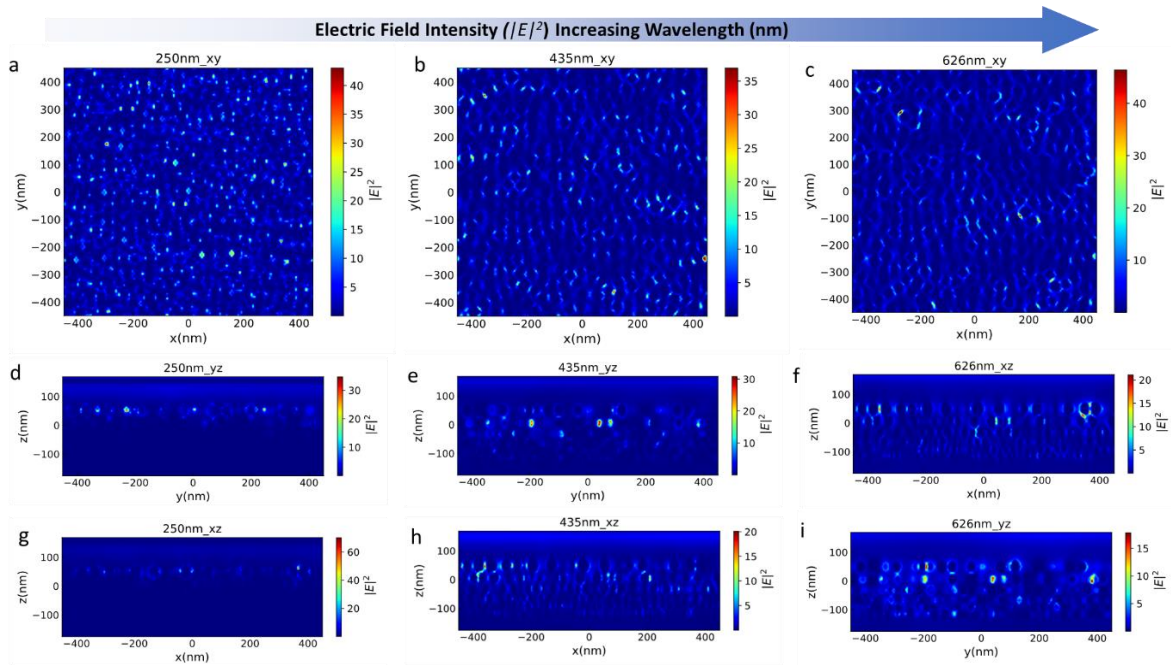

*Figure 26: Electric field intensity plots at UV (250nm) and visible (435nm and 626nm) regions for the six-Layered structure of PDMS 15. (a-c) The Intensity plots in the XY plane of the first layer of Ga nanodroplets. (d-f) Field intensity at YZ plane. (g-i) Field intensity at XZ plane.*

The following plot depicts the electric field intensity at some localized regions between two different layers of the Gallium nanodroplets. The incident electric field effectively sees a structure of feature size of the order of depth of the excited layers. Therefore the system in which the total thickness of Gallium droplet layers is larger will exhibit spectral features at larger wavelengths.

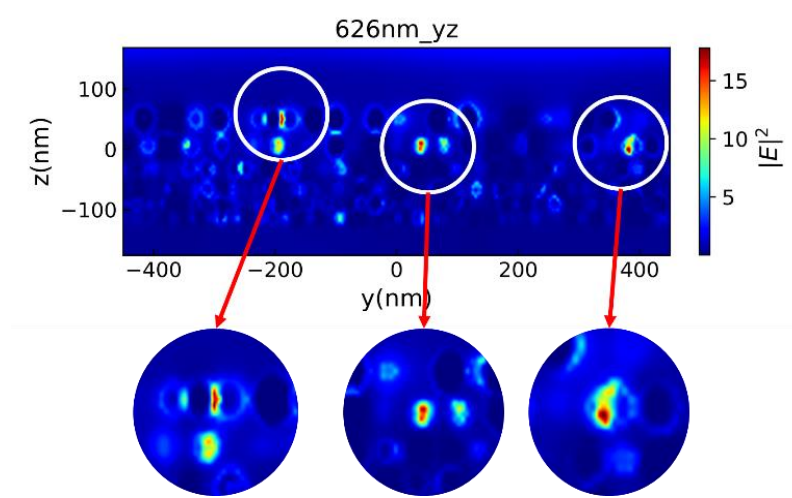

*Figure 27: Electric field intensity plots at some localized regions between two different layers of the Gallium nanodroplets. The inter-layer field interaction causes the light to interact with the droplets in all the layers, thus equivalent to the interaction of light with an object of effective length scale equal to the thickness of the Ga embedded region.*

The SEM image shown below reveals that the depth of the Ga nanodroplets is higher in PDMS 20. Since it has more layers of Ga nanodroplets, gap plasmon contribution from all the layers of PDMS 20 will occur at a wavelength higher than that in the case of PDMS 5. Therefore, the major spectral features of PDMS 20 are expected to be red-shifted compared to PDMS 5.

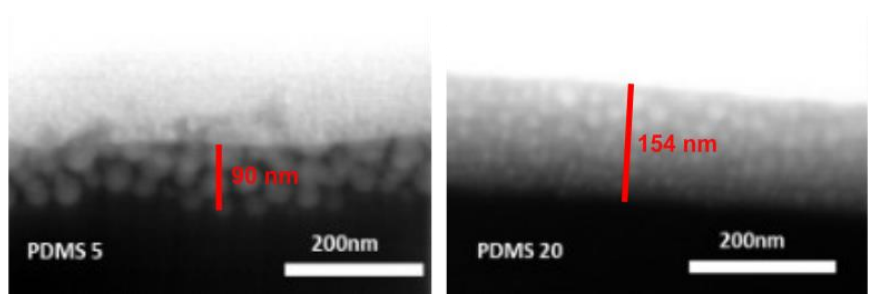

*Figure 28: The depth of Ga nanodroplets in the case of PDMS 20 (more oligomer content) is more than that of PDMS 5 (less oligomer content).*

**3.2 Mechanoresponsive on stretching the sample (uniaxial linear stretch)**

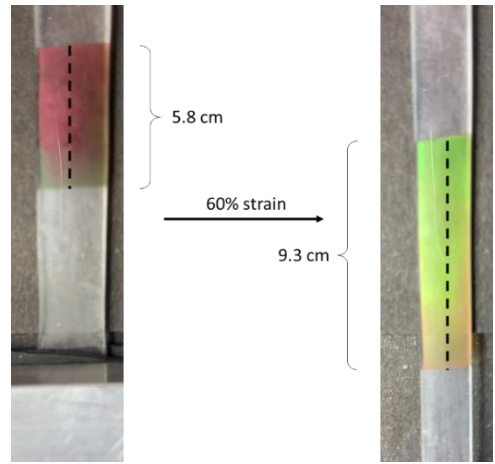

*Figure 29: Change of color of Ga deposited PDMS, occurring due to a strain of 60%.*

The change of color of the sample is attributed to the change in the chromogenic structure of Ga nanoparticles interacting with the incident light. For two adjacent particles in a given layer, stretching will increase the inter-droplet gap filled with PDMS oligomers with a refractive index of 1.4. Hence, the study is necessary to determine the interaction of electromagnetic fields with two Ga droplets in PDMS placed near each other with varying gaps.

**3.3 Study of two droplets of the same radius in an environment of PDMS**

For finite difference time domain (FDTD) simulations, the commercial software package Lumerical is used. The two spherical spheres are placed close to each other with a particular gap inside the FDTD region. Perfectly Matched Layer Boundary condition is applied on all sides. The source used is Total Field Scattered Field (TFSF), and the reflectivity monitor is placed outside the TFSF region behind the source.

In this section, we will show the following:

### 3.3.1 For polarisation vector along the Gap between the spheres

1) The fluidic interactions of liquid oligomers result in Ga nanodroplets of a radius of no more than 50nm. When two Ga nanospheres of the same radius (< 50nm) are embedded in PDMS (dielectric of refractive index 1.4), they exhibit gap plasmon resonance in the visible range if the polarisation vector is along the Gap between the spheres. It is the primary cause of resonance in this configuration.
2) The gap plasmon resonance blue shifts with the increase in the Gap (Figure 24a).
3) The resonances at shorter wavelengths correspond to collective dipole and Quadrupole oscillations of electrons in both spheres.
4) With the decrease in radii, the plasmon resonance blue shifts (Figure 25b).

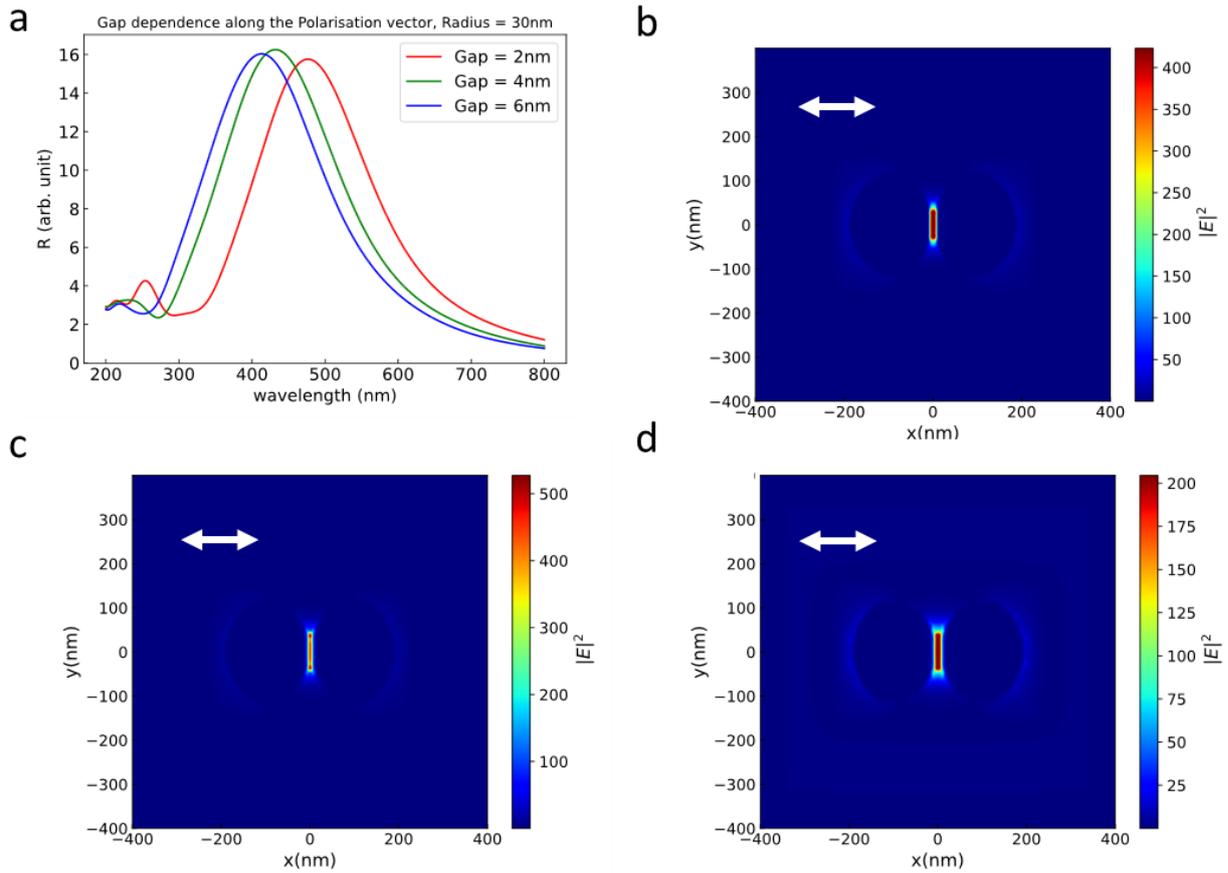

Figure 30: Gap plasmons variation with dielectric Gap. a) Reflectivity plot obtained from FDTD simulation. It indicates the blue-shift of the spectrum with the increase in the Gap between the two spheres. The Gap plasmon resonance occurs in the visible region. b-d) The field plots plotted at the gap plasmon resonance for a gap of 2, 4 and 6 nm respectively.

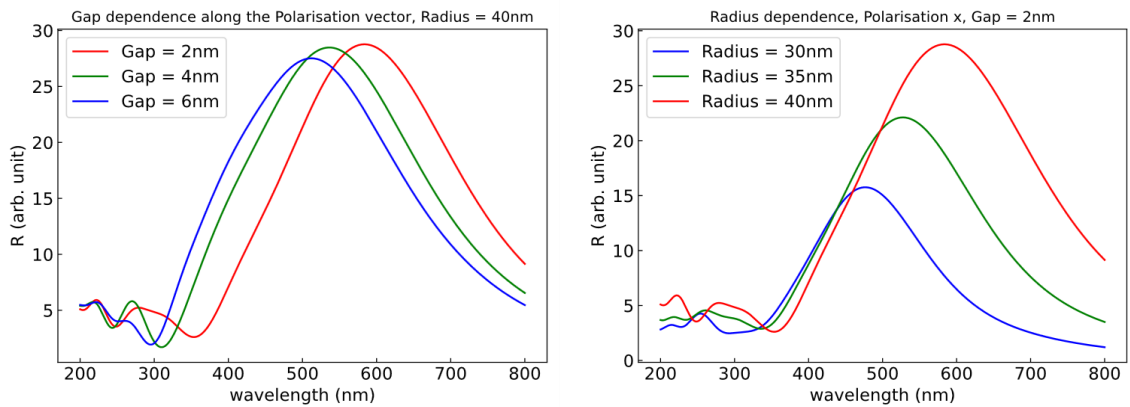

Figure 31: Reflectivity plot from two spheres at the different gap (left) and different radii(right).

## 3.3.2 For polarisation vector perpendicular to the gap between the spheres

1) The polarization vector perpendicular to the gap shows negligible interaction between the fields scattered by two nanospheres. This phenomenon is effectively a single-particle effect and can be approximated as two independent single-particle scattering events co-occurring, as indicated by the simulation results in Figure 26.
2) Since these scatterings are independent and the Ga droplets are spherical, we can apply Mie theory to understand the spectral features in the UV region.
3) The primary cause of resonance is the collective oscillation of electrons in the dipole mode of the spheres.
4) The reflectivity spectra of these systems of spheres are independent of the gap between them.

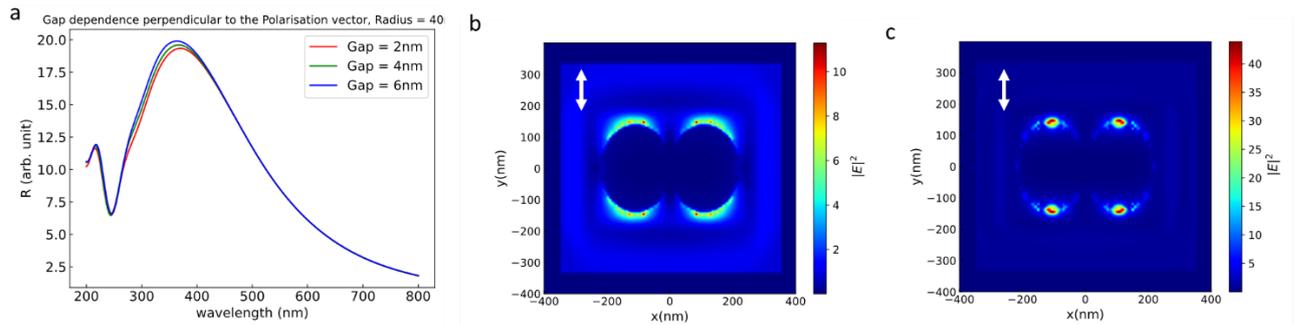

*Figure 32: Independent scattering by the two nanospheres in the case where polarisation is perpendicular to the direction of the gap. a) The reflectivity spectrum does not change by increasing the gap between the spheres. The major resonance is due to dipole resonance at the UV region, as plotted in (b). The minor peak is due to quadrupole resonance at the extreme UV region, as shown in (c).*

The above simulation readily predicts the following spectral behavior observed experimentally on applying a uniaxial strain to the sample. We observe that the spectral features in the UV region do not change, and those in the visible and IR region get blue-shifted (Figure 26a).

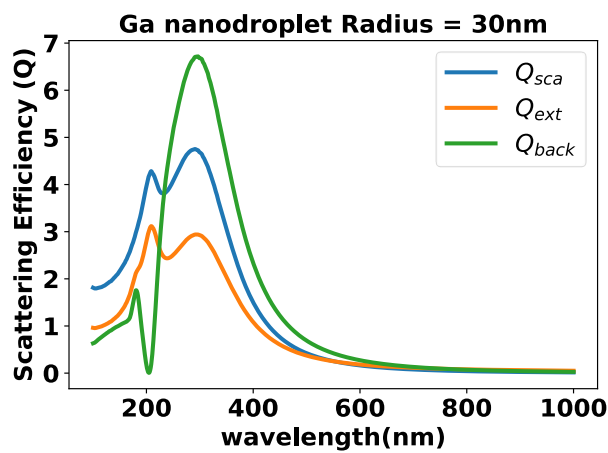

*Figure 33: Efficiencies of Scattering, Extinction, and back-scattering of Ga sphere of 30 nm radius embedded in a medium of refractive index 1.4 as calculated from Mie Theory.*

The UV spectral features can be attributed to the independent single-particle scattering of light by the individual Ga nanodroplets resulting from the interaction of those electric field components perpendicular to the gap between particles. As shown previously with the two-particle simulation, the spectral features due to single-particle scattering do not exhibit significant change. As shown above, in Figure 27, the scattering by the individual Ga spheres is limited to the UV region. On the other hand, the spectral features in the visible and IR region get blue-shifted, whose cause can be attributed to the increase in the inter-particle gap due to the uniaxial stretching (Figure 26a and 28).

In the following, we show the blueshift trend in the visible region and the null shift in the UV region of reflectivity spectra of different samples with respect to uniaxial stretching (Figure 28). It is corroborated by multiple and distinct simulations exhibiting the aforementioned spectral trends on applying a uniaxial strain.

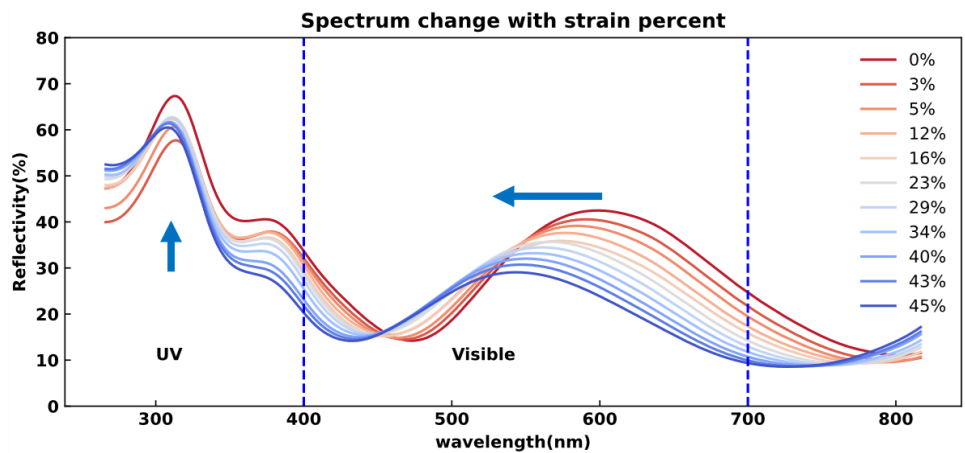

*Figure 34: Experimentally obtained spectra of Ga deposited PDMS sample (Thickness: 100nm, Temperature: 35C, PDMS 10) at different strain percentages. The blue shift occurs in the visible and IR regions, whereas no such shift is observed in the UV region.*

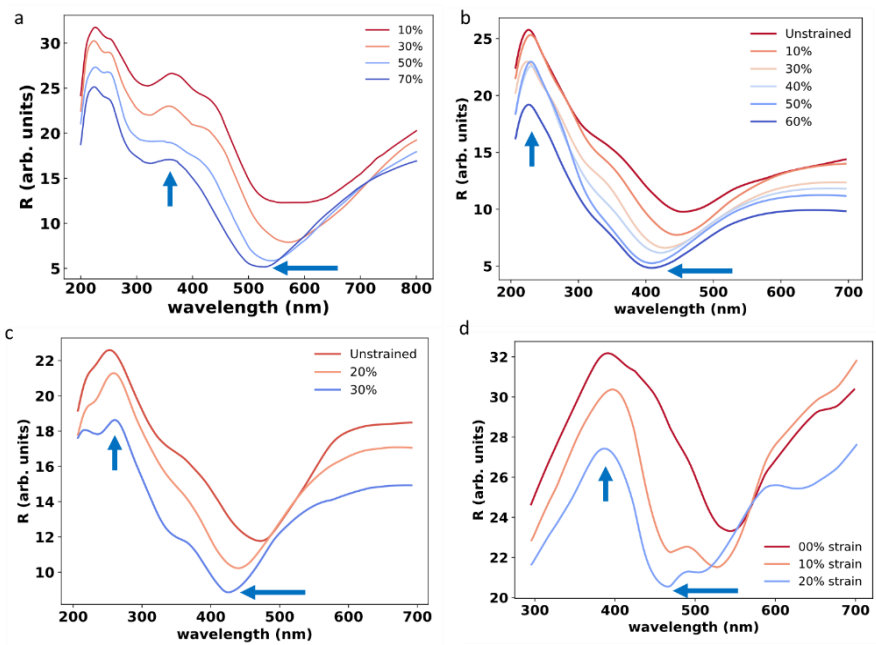

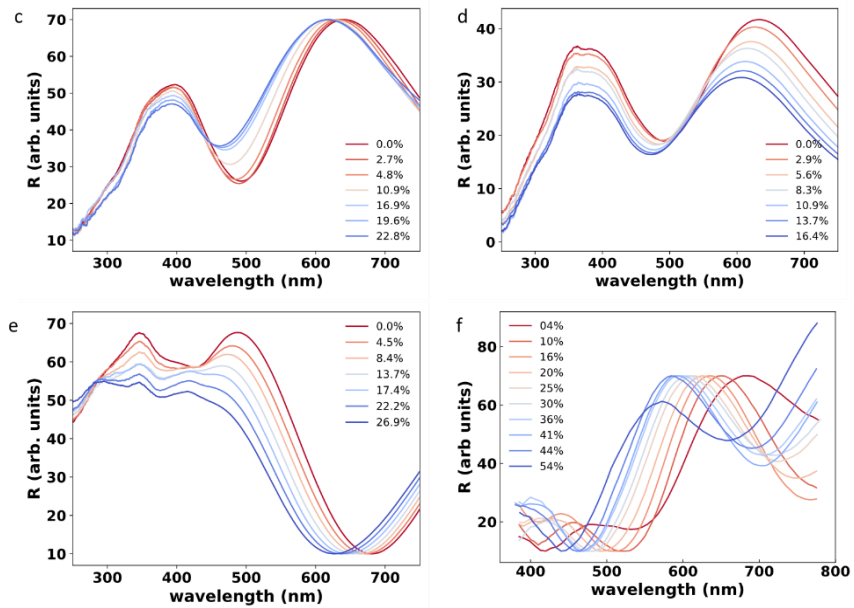

*Figure 35: Blue shift in the visible region and null-shift in the UV region of simulated and experimental spectra of Ga deposited PDMS sample at different strain percentage. a, b, c, d, Reflectivity spectra obtained from four different simulated structures varying in number of layers, and size distributions. e, f, g, h, Experimentally obtained reflectivity spectra of four different fabricated samples.*

# 4. Color characterization and other applications

**4.1 CIE coordinates from Reflectivity Spectra**

The following steps are followed to obtain the CIE x and y coordinates for a given reflectivity spectrum [6].

1. A reflectivity spectrum obtained in arbitrary units and normalized with a known reference (such as a silver mirror) is needed as an input.
2. We have the illuminant D65 spectrum as the source of light that determines illumination. The response of human eyes to different wavelengths is encoded in terms of three functions called the color-matching functions, denoted by $\bar{x}, \bar{y}$ and $\bar{z}$.

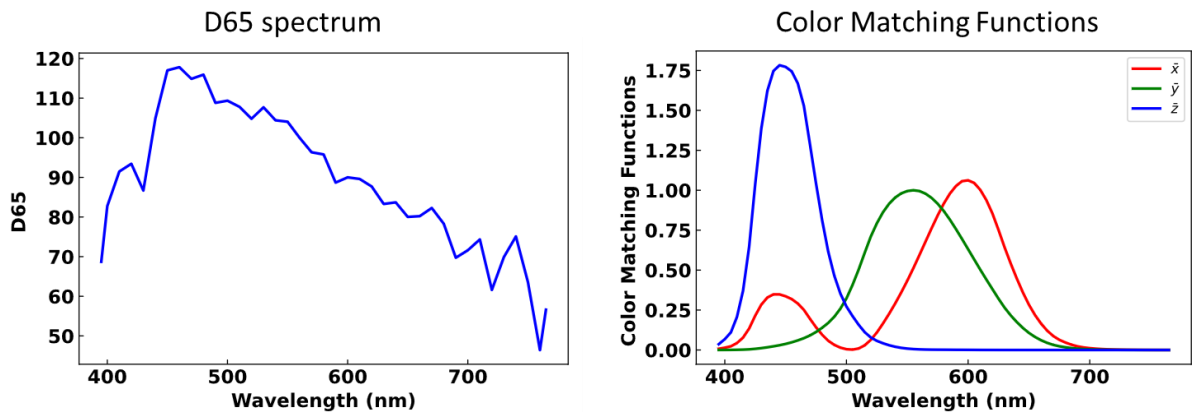

*Figure 36: Data that encapsulates the illuminant and response of eyes to a spectrum. a, Spectrum of illuminant D65. b, Color matching functions.*

3. The tristimulus for white light (D65) is obtained by,

$$X_{D65} = \frac{\sum \bar{x}.D65}{\sum \bar{y}.D65} = 0.950$$

$$Y_{D65} = \frac{\sum \bar{y}.D65}{\sum \bar{y}.D65} = 1.000$$

$$Z_{D65} = \frac{\sum \bar{z}.D65}{\sum \bar{y}.D65} = 1.088$$

where the dot (.) operation gives another array with each element as obtained from the elementwise multiplication of the two arrays. The expression $\sum \bar{x}.D65$ refers to the sum of the elements in the array obtained by elementwise multiplication of $\bar{x}$ and $D65$.

4. The white point is given by,

$$x_{D65} = \frac{X_{D65}}{X_{D65} + Y_{D65} + Z_{D65}} = 0.313$$

$$y_{D65} = \frac{Y_{D65}}{X_{D65} + Y_{D65} + Z_{D65}} = 0.329$$

5. Then we calculate $S$ by elementwise multiplication of the reflectivity spectrum and $D65$,

$$S = R \cdot D65$$

6. The tristimulus values are then obtained for $S$,

$$X = \frac{\sum \bar{x}.S}{\sum \bar{y}.D65}$$

$$Y = \frac{\sum \bar{y}.S}{\sum \bar{y}.D65}$$

$$Z = \frac{\sum \bar{z}.S}{\sum \bar{y}.D65}$$

7. The CIE x and y coordinates can be obtained from the tristimuli values as follows,

$$x = \frac{X}{X+Y+Z}$$

$$y = \frac{Y}{X+Y+Z}$$

The CIE chromaticity coordinates are given by $x$ and $y$, while the value of $Y$ denotes the object's brightness from which the reflectivity spectrum is measured.

**RGB and Hue from CIE xyY coordinates**

For a given color specified by CIE x, y and Y, one can obtain the RGB values by following steps:

1. We obtain $z$ by,

$$z = 1 - x - y$$

2. The rest of the tristimuli values are determined as,

$$X = \frac{xY}{y}$$

$$Z = \frac{zY}{y}$$

3. We obtain $R', G', and\ B'$ values from the following matrix operation,

$$\begin{pmatrix} R' \\ G' \\ B' \end{pmatrix} = \begin{pmatrix} 3.2404542 & -1.5371385 & -0.4985314 \\ -0.9692660 & 1.8760108 & 0.0415560 \\ 0.0556434 & -0.2040259 & 1.0572252 \end{pmatrix} \begin{pmatrix} X \\ Y \\ Z \end{pmatrix}$$

4. Let $S'$ and $S$ belong to the corresponding values in $(R', G', B')$ and $(R, G, B)$ respectively.

5. The values of $R, G$ and $B$ are obtained by the following conditional operation

   If $S' < 0.0031308$,
   $$S = 12.92\ S'$$
   Else,
   $$S = 1.055\ \ S'^{(1.0/2.4)} - 0.055$$

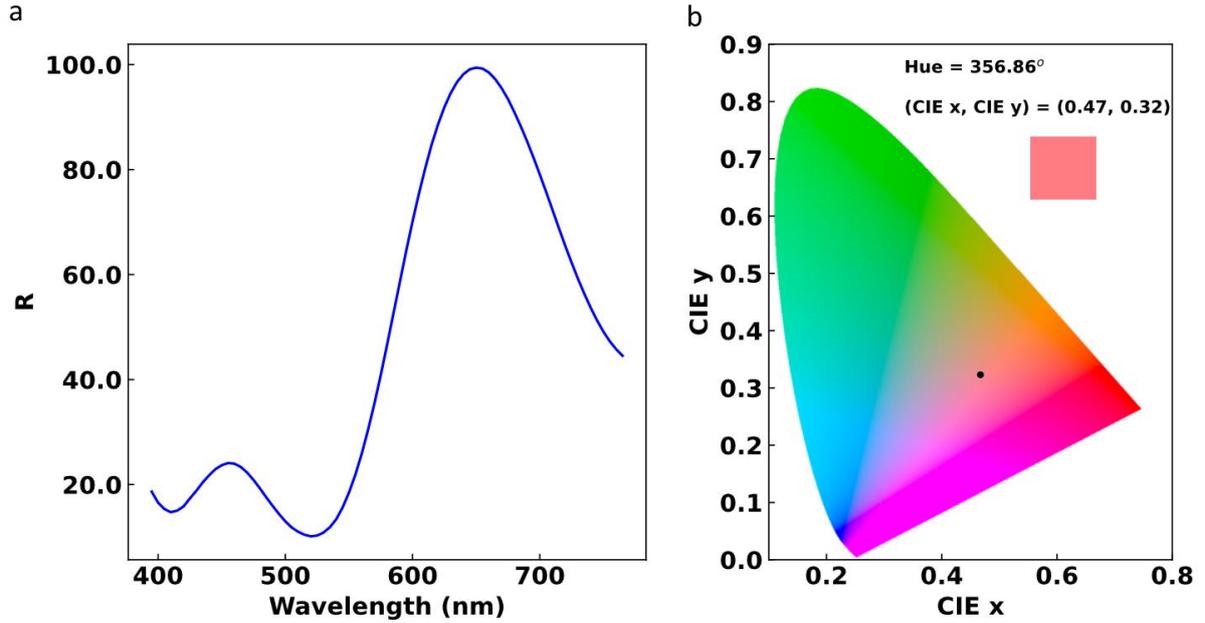

*Figure 37 Reflectivity plot and corresponding color coordinates. a, Reflectivity spectrum obtained from spectrophotometer and normalized with respect to the reference. b, The CIE x and y coordinates and hue corresponding to the reflectivity spectrum.*

6. Having obtained the values of $R, G, B$, we obtain hue by the following procedure:
   a. We evaluate $C_{max}, C_{min}$, and $\Delta$ as,
   $$C_{max} = max(R, G, B)$$
   $$C_{min} = min(R, G, B)$$
   $$\Delta = C_{max} - C_{min}$$
   b. If $\Delta = 0$, the value of the hue is $H = 0$
   c. If $C_{max} = R$, the value of the hue is $H = 60 \times \left(\frac{G-B}{\Delta}\right)$
   d. If $C_{max} = G$, the value of the hue is $H = 60 \times \left(\frac{B-R}{\Delta} + 2\right)$
   e. If $C_{max} = B$, the value of the hue is $H = 60 \times \left(\frac{R-G}{\Delta} + 4\right)$
7. The hue range obtained from the above procedure is from 0 to 360 degrees. In cases where the hue values are reported more than 360 degrees in the main text, the hue is to be wrapped to the range mentioned above by a modular division by 360. For example, hue value of 400 degrees is equivalent to 40 degrees (= 400 mod 360).

**4.2 Determination of Curvature from the image.**

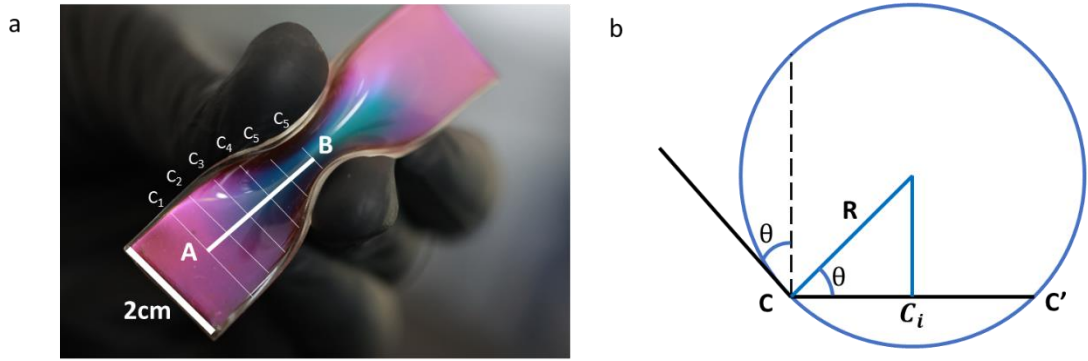

*Figure 38 Determination of the radius of curvature from the image along the line AB. a, Optical image of the sample showing a color change with varying curvature. b, The geometrical construction used to determine the radius of curvature for chord length $C_i$.*

To obtain the curvature from point A to B along the line AB we assume that linear variation of the edge angle from 0 to 90 degrees. The chord lengths $C_i$ (= CC') are determined to vary from 2cm to 0.47 cm at intermediate intervals along AB. The geometry (Figure 32b) is used to determine the curvature using the above data.

The corresponding hue can be computed by the abovementioned method after determining the RGB values using image processing software (ImageJ FIJI).

$$R = \frac{C_i}{2 \cos \theta}$$